\documentclass[aps,twocolumn,superscriptaddress,floatfix,longbibliography]{revtex4-1} %linenumbers

\usepackage{lipsum}
\usepackage{graphicx}
\graphicspath{{figures/Fig_Cartoon/}{figures/Fig_Clas_QM/}{figures/Fig_Diffusion/}{figures/Fig_Linearity/}{figures/Fig_Position/}{figures/Fig_Wavfunction/}}
\usepackage[export]{adjustbox}
\usepackage{amsmath,amssymb}
\usepackage{braket}
\usepackage{mathtools}
\usepackage{hyperref}
\usepackage{mathrsfs}
\usepackage{verbatim}
\usepackage{dsfont}

\usepackage{color}
\newcommand{\Z}{\mathbb{Z}}
\newcommand{\C}{\mathbb{C}}

\newcommand{\n}[1]{\left| #1 \right|}%%adjustable-height norm shortcut
\newcommand{\dn}[1]{\left|\left| #1 \right|\right|}%%adjustable-height norm shortcut
\newcommand{\st}[1]{\left\{#1\right\}}%%adjustable-height set notation

\renewcommand{\v}[1]{\boldsymbol{#1}}%%shortcut to make a vector (overwrites the default command)
\DeclareMathOperator{\Tr}{Tr}

\renewcommand{\O}{\mathcal{O}}
\newcommand{\BigO}{\mathrm{O}}
\newcommand{\smallo}{\mathrm{o}}
\def\pbra#1{\mathinner{(#1|}}
\def\pket#1{\mathinner{|#1)}}
\def\pbraket#1{\mathinner{\left(#1\right)}}

\def\L{\mathcal{L}}

\DeclareMathOperator{\sech}{sech}
\newcommand*\pFq[2]{{}_{#1}F_{#2}}

\begin{document}
	\title{A Universal Operator Growth Hypothesis}

	\author{Daniel E. Parker}
	\email[]{daniel\_parker@berkeley.edu}
	\affiliation{Department of Physics, University of California, Berkeley, CA 94720, USA}
	
	\author{Xiangyu Cao}
	\email[]{xiangyu.cao@berkeley.edu}
	\affiliation{Department of Physics, University of California, Berkeley, CA 94720, USA}
	
	\author{Alexander Avdoshkin}
	\email[]{alexander_avdoshkin@berkeley.edu}
	\affiliation{Department of Physics, University of California, Berkeley, CA 94720, USA}
		
	\author{Thomas Scaffidi}
	\email[]{thomas.scaffidi@berkeley.edu}
	\affiliation{Department of Physics, University of California, Berkeley, CA 94720, USA}
	\affiliation{Department of Physics, University of Toronto, Toronto, Ontario, M5S 1A7, Canada}
	
	\author{Ehud Altman}
	\email[]{ehud.altman@berkeley.edu}
	\affiliation{Department of Physics, University of California, Berkeley, CA 94720, USA}
	
	\date{\today}
	
	\begin{abstract}
		We present a hypothesis for the universal properties of operators evolving under Hamiltonian dynamics in many-body systems. The hypothesis states that successive Lanczos coefficients in the continued fraction expansion of the Green's functions grow linearly with rate $\alpha$ in generic systems, with an extra logarithmic correction in 1d. The rate $\alpha$ --- an experimental observable --- governs the exponential growth of operator complexity in a sense we make precise. This exponential growth prevails beyond semiclassical or large-$N$ limits. Moreover, $\alpha$ upper bounds a large class of operator complexity measures, including the out-of-time-order correlator. As a result, we obtain a sharp bound on Lyapunov exponents $\lambda_L \leq 2 \alpha$, which complements and improves the known universal low-temperature bound $\lambda_L \leq 2 \pi T$. We illustrate our results in paradigmatic examples such as non-integrable spin chains, the Sachdev-Ye-Kitaev model, and classical models. Finally we use the hypothesis in conjunction with the recursion method to develop a technique for computing diffusion constants.
	\end{abstract}
	\maketitle
	%except for 1d quantum systems, in which there is an extra logarithmic correction.
	
	\tableofcontents
	\section{Introduction}
	
	The emergence of ergodic behavior in quantum systems is an old puzzle. Quantum mechanical time-evolution is local and unitary, but many quantum systems are effectively described by irreversible hydrodynamics, involving familiar quantities such as electrical conductivity. Understanding this emergent thermal behavior at both a conceptual and computational level is a central goal of theoretical research on quantum dynamics, of which a cornerstone is the Eigenstate Thermalization Hypothesis~\cite{deutsch91,srednicki94,rigol2008thermalization,alessio16eth,deutsch2018eigenstate}. 
	
	Recent work has shifted focus from states to \textit{operator growth} in many-body systems~\cite{von2018operator, nahum2018operator, rakovszky2017diffusive,khemani2018operator,gopalakrishnan2018hydrodynamics,chan2018solution}. Under Heisenberg-picture evolution, simple operators generically decay into an infinite ``bath'' of increasingly non-local operators. The emergence of this dissipative behavior from unitary dynamics is believed to be at the origin of thermalization, the decay of dynamical correlation functions, and the accuracy of hydrodynamics at large scales.  This picture was recently confirmed in random unitary models of quantum dynamics~\cite{von2018operator, nahum2018operator}, and extended to increasingly realistic systems involving conservation laws~\cite{rakovszky2017diffusive,khemani2018operator}, Floquet dynamics~\cite{chan2018solution}, and even interacting integrable models~\cite{gopalakrishnan2018hydrodynamics}.
	
	While random unitary models are valuable proxies for studying operator growth, one would like to confirm this picture in genuine Hamiltonian systems. In semiclassical systems, a quantitative measure is provided by the out-of-time-order correlation function (OTOC). The classical butterfly effect gives rise to an exponential growth of the OTOC, characterized by the Lyapunov exponent, which may be computed in a variety of models. It is conjectured that the Lyapunov exponent is bounded~\cite{maldacena2016bound} and this bound is achieved in certain large-$N$ strongly interacting models with a classical gravity dual, such as the Sachdev-Ye-Kitaev (SYK) model~\cite{sykcomment,kitaev15,SY93}. Unfortunately, the OTOC does not necessarily exhibit exponential growth outside of semiclassical or large-$N$ limits, rendering the Lyapunov exponent ill-defined~\cite{elsayed14,xu2018accessing,khemani2018operator,xu2018locality}. A general theory of operator growth under generic, non-integrable Hamiltonian dynamics is, therefore, still lacking. 
	
	The amount of information required to describe a growing operator increases exponentially in time. Computationally, this bars the exact calculation of operators at long times. Yet, the exponential size of the problem has a positive aspect: it acts as a thermodynamic bath, so a statistical description should emerge and become nigh-exact. This idea indicates operator growth should be governed by some form of universality.  In this work we present a hypothesis specifying universal properties of growing operators in non-integrable quantum systems in any dimension.
	
	\section{Synopsis}
	Our hypothesis has a simple formulation in the framework of the continued fraction expansion or \textit{recursion method}, which we review in Section~\ref{sec:review}. This is a well-understood technique, dating back to the 1980s~\cite{mattis1981reduce}, and has recently been used to compute conductivities in strongly-interacting systems \cite{lindner2010conductivity,khait16mbl,auerbach2018hall}. It is surveyed in great detail in Ref. \cite{viswanath2008recursion}.  Essentially, it converts any linear-response calculation to the problem of a quantum  particle on a half chain, with the hopping matrix elements given by the Lanczos coefficients $b_n$.
	Section~\ref{sec:hypo} presents our hypothesis: operators in generic, non-integrable systems have Lanczos coefficients with asymptotically linear growth with $n$, suppressed by a logarithmic correction in one dimension. The linear growth rate, denoted $\alpha$, is the central quantity of this work. It has dimensions of energy and can be bounded from above by the local bandwidth [see \eqref{eq:alphabound} and \eqref{eq:alphaboundwithlog}]. In light of this, the hypothesis essentially asserts that the Lanczos coefficients grow as fast as possible in non-integrable systems. Although we are unable to prove the hypothesis rigorously, we shall support it with extensive numerical evidence, calculations in SYK models, and general physical arguments in Section~\ref{sec:hypo}. In particular, the hypothesis is equivalent to the exponential decay of the spectral function at high frequency, which can be (and has been) observed experimentally~\cite{PhysRev.188.609,PhysRevB.10.822,0953-8984-2-50-017}. 
	
	We explore several consequences of the hypothesis. In Section~\ref{sec:complexity}, we develop a precise picture of the universal growth of operators. We show that under the hypothesis, the 1d quantum mechanics, governed by the Lanczos coefficients $b_n \sim \alpha n$, captures the irreversible process of simple operators evolving into complex ones. Furthermore, the  1d wavefunction delocalizes exponentially fast on the  $n$ axis, at a rate exactly given by $\alpha$. Asymptotically, the expected position of the 1d wavefunction satisfies
	\begin{equation} \label{eq:expected_position_intro}
	(n)_t \sim e^{2 \alpha t} \,.
	\end{equation} The expectation value $\pbraket{n}_t$ has a succinct interpretation as an upper bound for a large class of operator complexity measures  called ``q-complexities'', which we define in section \ref{sec:complexity_bound}.  Crucially, this class includes out-of-time-order correlators. This allows us to establish a quantitative connection between $\alpha$ and the Lyapunov exponent, which will be the subject of Section~\ref{sec:chaos_bound}. We show for quantum systems at infinite temperature that the growth rate gives an upper bound for the Lyapunov exponent whenever the latter is well-defined:
	\begin{equation} \label{eq:Lyapunov_conj_intro}
	\lambda_L \leq 2 \alpha \,.
	\end{equation}
	For classical systems, this statement is a conjecture but it is posible to prove a somewhat weaker bound. We check \eqref{eq:Lyapunov_conj_intro} in the SYK model and a classical tops model, and find it to be tight in both cases. 
	
	A further application of the hypothesis, discussed in Section~\ref{sec:decay}, is a semi-analytical technique to compute diffusion coefficients of conserved quantities. We leverage the hypothesis to extend classical {methods} of the continued fraction expansion to directly compute the pole structure of the Green's function, thus revealing the dispersion relation of the dynamics. 
	
   Section~\ref{sec:finite_temp} discusses the generalization to finite temperatures, which involves many open questions. Nevertheless, we show that the universal bound on chaos $\lambda_L \le 2 \pi k_B T/\hbar$~\cite{maldacena2016bound} can be implied and improved by a proper finite-temperature extension of the bound \eqref{eq:Lyapunov_conj_intro}, and provide evidence supporting this conjecture. We conclude in Section~\ref{sec:discussion} by discussing conceptual implications of our results and perspectives for future work. 
	%\Thomas{do we really need the qualifier ``probably'' here?}\
	
	\section{Preliminaries: The Recursion Method}
	\label{sec:review}
	
	We briefly review the recursion method in order to state the hypothesis. A comprehensive treatment may be found in \cite{viswanath2008recursion}. Consider a local Hamiltonian $H$ and fix a Hermitian operator $\O$. We regard the operator as a state $\pket{\O}$ in the Hilbert space of operators, endowed with the infinite-temperature inner product $\pbraket{\O_1|\O_2} := \Tr[\O_1^\dagger \O_2] / \Tr[1]$. We write $\dn{\O} := \pbraket{\O|\O}^{1/2}$ for the norm. We will focus on systems in the thermodynamic limit. 
	
	Just as states evolve under the Hamiltonian operator, operators evolve under the Liouvillian superoperator $\L := [H,\cdot]$. Our central object is the autocorrelation function
	\begin{equation}
	C(t) = \Tr[\O(0) \O(t)] / \Tr[1] = \pbraket{\O|\exp\left(i \L t \right) | \O},
	\label{eq:autocorrelation}
	\end{equation}
	where the second equality follows from Baker-Campbell-Hausdorff. 
	
	%One is interested in the long-time behavior of $C(t)$, and also its frequency dependence ({encoded in the} spectral function).

\begin{figure}[h]
		\centering
		\includegraphics[width=\linewidth]{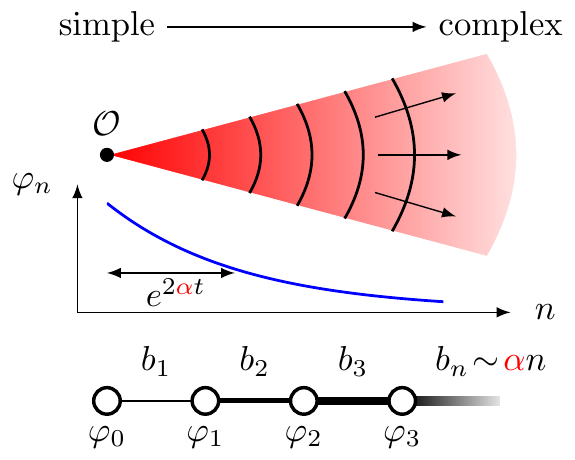}
		\caption{Artist's impression of the space of operators and its relation to the 1d chain defined by the Lanczos algorithm starting from a simple operator $\mathcal{O}$. The region of complex operators corresponds to that of large $n$ on the 1d chain. Under our hypothesis, the hopping amplitudes $b_n$ on the chain grow linearly asymptotically in generic thermalizing systems (with a log-correction in one dimension, see Section~\ref{sec:1dspecial}). This implies an exponential spreading $\pbraket{n}_t \sim e^{2\alpha t}$ of the wavefunction $\varphi_n$ on the 1d chain, which reflects the exponential growth of operator complexity under Heisenberg evolution, in a sense we make precise in Section~\ref{sec:complexity}.  The form of the wavefunction $\varphi_n$ is only a sketch; see Fig.~\ref{fig:exponential_spreading_wavefunction} for a realistic picture.}
		\label{fig:Liouvillian_graph}
	\end{figure}

	\begin{comment}
	There are several equivalent ways to reexpress the autocorrelation function in common usage. The Green's function is defined via
	\begin{equation}
		G(z) = \pbraket{\O\left|\frac{1}{z - \L}\right|\O} = i \int_0^\infty e^{-izt} C(t)\, dt.
		\label{eq:Green_fcn_defn}
	\end{equation}
	Alternatively, the spectral function is given by
	\begin{equation}
		\begin{aligned}
		\Phi(\omega) \ &=\ \int_{-\infty}^\infty C(t) e^{-i\omega t} \, dt\\
		\ &=\ \sum_{E,E'} \n{\braket{E|\O|E'}}^2 \delta(\omega-E'+E).\\
	\end{aligned}
		\label{eq:spectral_fcn_defn}
	\end{equation}
	Each of these is a linear transformation of the correlation function, and highlights different aspects of the dynamics: the dispersion relation from the Green's function and the distribution of eigenstates from the spectral representation. The primary tool of this work is yet-another description of the dynamics that --- while less-commonly used --- reveals the thermal structure of the dynamics. The idea, originally due to Mattis, is to use a \textit{non-linear} transformation of the initial data to put the Liouvillian into tridiagonal form.
	
	Before recalling the details, we first provide a brief motivation. 
	\end{comment}
	Computing  $C(t)$  is inherently difficult. Suppose $\O(t=0)$ is a relatively simple operator that can be written as the sum of a few basis vectors in any local basis \footnote{A local basis in, say, a spin chain is any basis related to the basis of Pauli strings by a finite-depth local unitary circuit.}. As the spatial support of $\O(t)$ grows, the number of non-zero coefficients of $\O(t)$ in any local basis can blow up exponentially. To make progress, one must compress this information. Intuitively, there are so many basis vectors at a given spatial size or ``complexity'' that we can think of them as a thermodynamic bath; no single basis vector has much individual relevance, only their statistical properties are important. In this interpretation, the operator flows though a series of ``operator baths'' of increasing size. The dynamics of an operator is then reduced to how the baths are connected --- a much simpler problem. In particular, the second law then dictates that an operator eventually flows to the largest possible baths, running irreversibly away from small operators. This is shown schematically in Fig. \ref{fig:Liouvillian_graph}.
	
	We now quantify this idea precisely. This is done by applying the Lanczos algorithm, which iteratively computes a tridiagonal representation of a matrix. The idea is to find the sequence $\st{\L^n \pket{\O}}$, and then apply Gram-Schmidt to orthogonalize. Explicitly, start with a normalized vector $\pket{\O_0} := \pket{\O}$. As a base case, let $\pket{\O_1} := b_1^{-1} \L \pket{O_0}$ where $b_1 := \pbraket{\O_0 \L|\L \O_0}^{1/2}$. Then inductively define
	\begin{equation}
	\begin{aligned}
	\pket{A_n} &:=\L  \pket{ \O_{n-1}} - b_{n-1} \pket{ \O_{n-2}} \,, \\
	b_n &:= \pbraket{A_n \vert A_n}^{1/2} \,, \\
	\pket{\O_{n}} \ &:= \ b_{n}^{-1} \pket{A_n} \,. \\
	\end{aligned}
	\label{eq:Lanczos_algorithm}
	\end{equation}
	The output of the algorithm is a sequence of positive numbers, $\st{b_n}$, called the \textit{Lanczos coefficients}, and an orthonormal sequence of operators, {$\st{\pket{\O_n}}$}, called the \textit{Krylov basis}. (This is a bit of a misnomer, as the Krylov basis spans \textit{an} operator space containing $\O(t)$ for any $t$, but does not usually span the full space of operators). The Liouvillian is tridiagonal in this basis: 
	\begin{equation}
	L_{nm}  := \pbraket{\O_n\vert\L\vert\O_m } = \begin{pmatrix}
	0 & b_1 & 0 & 0 & \cdots\\
	b_1 & 0 & b_2 & 0 & \cdots\\
	0 & b_2 & 0 & b_3 & \cdots\\[-0.3em]
	0 & 0 & b_3 & 0 & \ddots\\[-0.2em]
	\vdots & \vdots & \vdots & \ddots & \ddots
	\end{pmatrix}.
	\label{eq:L_tridiagonal}
	\end{equation}
	
	We make four remarks. First, if the operator Hilbert space is $d$-dimensional with $d$ finite (or if the subspace spanned by  $\pket{\O_0}, \pket{\O_1}, \pket{\O_2}, \dots$ is so), the algorithm will halt at $n = d + 1$: in this work, we work always in the thermodynamic limit and discard this non-generic situation. Second, the Lanczos algorithm presented here is adapted to operator dynamics. Generally, a tridiagonal matrix will have non-zero diagonal entries, but they vanish in \eqref{eq:L_tridiagonal}. This is because one can inductively show that $i^n \O_n$ is Hermitian for all $n$, hence $\pbraket{\O_n|\L \vert \O_n} = 0$. Third, the knowledge of the Lanczos coefficients $b_1, \dots,b_n$ is equivalent to that of the \textit{moments} $\mu_2, \mu_4, \dots, \mu_{2n}$, defined as the Taylor series coefficients of the correlation function
	\begin{equation}
		\mu_{2n} := \pbraket{\O \vert \L^{2n} \vert \O} = \frac{d^{2n}}{dt^{2n}} C(t)\big|_{t=0}
	\end{equation}
	The non-trivial transformation between the Lanczos coefficients and the moments is reviewed in Appendix \ref{app:recursion-review}. Fourth, the Lanczos coefficients have units of energy.
	
	In the Krylov basis, the correlation function $C(t)$ is:
	\begin{equation}
	C(t) = \left(e^{iLt}\right)_{00} \,.
	\end{equation}
	Hence the autocorrelation depends only on the Lanczos coefficients, and not on the Krylov basis. One way to interpret the Lanczos coefficients, which we will employ extensively below, is as the hopping amplitudes of a semi-infinite tight-binding model --- see Fig. \ref{fig:Liouvillian_graph}. The wavefunction on the semi-infinite chain is defined as $\varphi_n(t) := i^{-n} \pbraket{\O_n|\O(t)}$. Heisenberg evolution {of $\O(t)$} becomes a discrete Schr{\"o}dinger equation:
	\begin{equation}
	\partial_t \varphi_n = -b_{n+1} \varphi_{n+1} + b_n \varphi_{n-1}, \quad \varphi_n(0) = \delta_{n0}.
	\label{eq:1D_chain_problem}
	\end{equation}
	where $b_0 = \varphi_{-1} = 0$ by convention.  The autocorrelation is simply $C(t) = \varphi_0(t)$, so the Lanczos coefficients are completely equivalent to the autocorrelation function.
	 
Just as different bases are well-suited for particular computations, a number of equivalent representions of the autocorrelation function appear in this work, namely the Green's function 
	\begin{equation}
		G(z) = \pbraket{\O\left|\frac{1}{z - \L}\right|\O} = i \int_0^\infty e^{-izt} C(t)\, dt.
		\label{eq:Green_fcn_defn}
	\end{equation}
	and the spectral function
	\begin{equation}
		\begin{aligned}
		\Phi(\omega) \ &=\ \int_{-\infty}^\infty C(t) e^{-i\omega t} \, dt \,.
		%\\ \ &=\ \frac1{\mathrm{Tr}[1]} \sum_{E,E'} \n{\braket{E|\O|E'}}^2 \delta(\omega-E'+E).\\
	\end{aligned}
		\label{eq:spectral_fcn_defn}
	\end{equation}
%(For the the spectral representation to make sense, we need to consider finite systems at first)
In summary, we have reviewed five equivalent ways to describe the dynamics
	\begin{equation}
		C(t) \leftrightarrow G(z) \leftrightarrow \Phi(\omega) \leftrightarrow \{\mu_{2n}\} \leftrightarrow \{b_n\}
		\label{eq:equiv_dynamics}
	\end{equation}
Just as with a choice of basis, we shall use the most convenient representation for the task at hand and translate freely between them. We note that $\{b_n\}$ is special in the sense that it is a non-linear representation of the autocorrelation while all other representations are linearly related. We provide the details on the mapping to $b_n$ in Appendix \ref{app:recursion-review}, with a particular focus on asymptotic properties. 
    
The nonlinearity involved in $\{b_n\}$ also makes them more abstract. Intuitively, we can think of the Krylov basis $\{\O_n\}$ as stratifying operators by their `complexity' (with respect to the initial operator $\O$), {and $b_n$'s describe how operators of different complexities transform into one another.}  The goal of this work is to study aspects of operator growth that can be reduced to the quantum mechanics on this semi-infinite chain. 

    % \sout{In terms of the bath picture above, the ``occupation number'' of the $n$th bath is $|\varphi_n|^2$, and the $b_n$'s describe the connections between adjacent baths } \Ehud{[I am not sure that the last sentece actually gives good intuition or is just confusing. The bath, if anything, is the rest of the chain from $n$ onward. It's probably not good to think of a single quantum state, the $n^{\text{th}]}$ basis state as a bath.]}. \Xiangyu{[The intuition we wanted to convey is that each site is the compression of a bath of many Pauli strings. But maybe it's probably safer to leave the sentence out.]}

	\section{The Hypothesis }
	\label{sec:hypo}

	We now state the hypothesis. Informally, \textit{in a chaotic quantum system, the Lanczos coefficients $\{b_n\}$ should grow as fast as possible}. The maximal possible growth rate turns out to be linear (with logarithm corrections in 1D). Our precise statement is therefore as follows. Suppose that $H$ describes an infinite, non-integrable~\footnote{As a working definition, we say that a system is integrable if it has an extensive number of quasi-local conserved quantities.}, many-body system in dimension $d>1$ and $\O$ is a local operator having zero overlap with any conserved quantity (in particular, $\pbraket{\O|H} = 0$). Then the Lanczos coefficients are asymptotically linear:
	\begin{equation}
	b_n = \alpha n + \gamma + \smallo(1),
	\label{eq:linear_lanczos}
	\end{equation}
	for some real constants $\alpha > 0$ and $\gamma$. This linear growth is an example of universality. We will refer to $\alpha$ as the \textit{growth rate}, and it will play a multitude of roles. In fact, it quantitatively captures the growth of ``operator complexity'' in a precise sense (Section~\ref{sec:complexity_bound}). On the other hand, it is observable by standard linear response measures (Section~\ref{sec:upper_bounds}). This section first describes why linear growth is maximal, amasses a weight of evidence in favor of the hypothesis, and finally discusses the special case of one dimension.

\begin{figure}[h]
		\centering
		\includegraphics{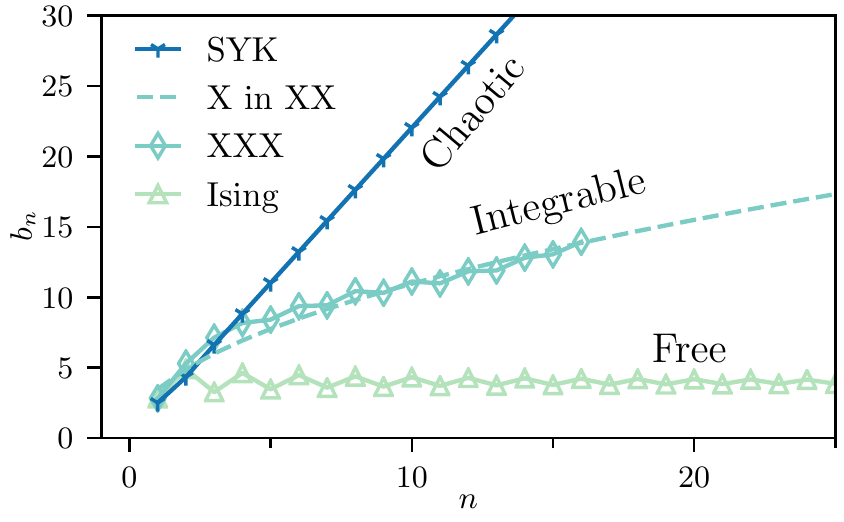}
		\caption{Lanczos coefficients in a variety of models demonstrating common asymptotic behaviors. ``Ising'' is $H = \sum_i X_i X_{i+1} + Z_i$ with $\O = \sum_j  e^{iq_j} Z_j$ ($q = 1/128$ here and below) and has $b_n \sim{} O(1)$. ``X in XX'' is $H = \sum_i X_i X_{i+1} + Y_i Y_{i+1}$ with $\O = \sum_j X_j$, which is a string rather than a bilinear in the Majorana fermion representation, so this is effectively an interacting integrable model that has $b_n \sim{} \sqrt{n}$. XXX is $H = \sum_i X_i X_{i+1} + Y_i Y_{i+1} + Z_i Z_{i+1}$ with $\O = \sum_j  e^{iq_j} (X_j Y_{j+1} - Y_j X_{j+1})$ that appears to obey $b_n \sim{} \sqrt{n}$. Finally, SYK is ~\eqref{eq:Q_SYK_model} where $q=4$ and $J= 1$ and $\O = \sqrt{2}\gamma_1$ with $b_n \sim{} n$. The Lanczos coefficients have been rescaled vertically for display purposes. Numerical details are given in Appendices~\ref{app:SYK} and ~\ref{app:numerics}.
		}
		\label{fig:patterns}
	\end{figure}

	We note that the idea of classifying operator dynamics by Lanczos coefficients asymptotics is as old as the recursion method itself. Many examples have been explored, resulting in a broad zoology, as surveyed in~\cite{viswanath2008recursion}. In particular, it is known that non-interacting models (such as lattice free fermions) give rise to a \textit{bounded} sequence $b_n \sim{} \BigO(1)$. If we start with a two-body operator $\O$ in such free models, all $\O_n$'s will remain two-body. In this sense, the operator dynamics is simple. In this work, we focus on the opposite extreme of generic chaotic dynamics. To our knowledge, the ubiquity of asymptotically linear growth in these systems and its consequences have not been systematically studied in quantum systems. Interacting models with obstructions to thermalization (e.g., integrable systems) lead to more involved behaviors, which have not been thoroughly explored. Nevertheless, a square root behavior $b_n \sim \sqrt{n}$ has been observed in a few examples~ (\cite{viswanath2008recursion,lee2001ergodic}, see also Fig.~\ref{fig:patterns}).
	
\subsection{Upper Bounds}
\label{sec:upper_bounds}

We start by showing that linear growth is the maximal possible growth of the Lanczos coefficients. This is most easily done starting with the spectral function. In interacting many-body systems, the spectral function has a tail extending to arbitrarily high frequencies. The asymptotic behavior of the tail is directly related to the Lanczos coefficients, with faster growth of Lanczos coefficients corresponding to slower decay of $\Phi(\omega)$. The precise asymptotic behavior is~\cite{Lubinsky1987,pettifor2012recursion}
	\begin{equation} 
	b_n \sim n^{\delta} 
	\Longleftrightarrow
	\Phi(\omega) \sim \exp(- |\omega/\omega_0|^{1/\delta}) \,    
	\label{eq:magnus_theorem}
	\end{equation}
	for any $\delta > 0$ and some constant $\omega_0$. In particular, $\delta = 1$ corresponds to asymptotically linear Lanczos coefficients and an exponentially decaying spectral function. 
	
 	The decay of the spectral function is constrained by a powerful bound. A rigorous and general result of Refs~\cite{abanin15} (see also~\cite{doublon,arad16,abanin17}, and Appendix~\ref{app:geometric_bound_1d} for a self-contained proof) is that, given an $r$-local lattice Hamiltonian $H = \sum_i h_i$ in any dimension,
	\begin{equation}
	\Phi(\omega) \le C e^{-\kappa |\omega|},\; \kappa = \frac{1}{2e G_r ||h_i||}
	\label{eq:spectral_fcn_bound}
	\end{equation}
	for some $C>0$ and a known $\BigO(1)$ geometrical factor $G_r$. We may conclude $\delta \le 1$ in \eqref{eq:magnus_theorem}, so the Lanczos coefficients grow at most linearly. 

\begin{figure}
		\centering
		\includegraphics[width=.75\columnwidth]{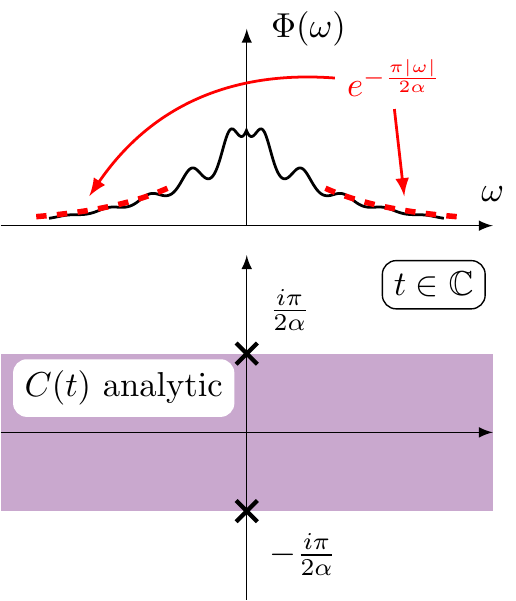}
		\caption{Illustration of the spectral function and the analytical structure of $C(t), t \in \C$. When the Lanczos coefficients have linear growth rate $\alpha$, $\Phi(\omega)$ has exponential tails $\sim e^{-|\omega|/\omega_0}$ with $\omega_0 = 2 \alpha / \pi $; $C(t)$ is analytical in a strip of half-width $ 1/\omega_0$ and the singularities  closest to the origin are at $t = \pm i / \omega_0$. See Appendix~\ref{app:recursion_2} for further discussion.} 
		\label{fig:strip}
	\end{figure}

	When linear growth of the $b_n$'s is achieved, the growth rate $\alpha$ is quantitatively related to the exponential decay rate in the spectral function. In fact, Appendix \ref{app:recursion-review} shows the following asymptotics are equivalent (see Fig.~\ref{fig:strip}):
	\begin{subequations} \label{eq:spectral_density_asymptotic}
		\begin{align}
		b_n &= \alpha\, n + \BigO(1) \,,\,   \\
		\Phi(\omega) &= e^{- \frac{|\omega|}{\omega_0} + \smallo(\omega)},\; \omega_0 = \frac2\pi \alpha,
		%    \mu_{2n}  &= \left( \frac{4 n \alpha}{e \pi} \right)^{2n} e^{\smallo(n)}.
		\end{align}
	\end{subequations}
	We stress that this is a purely mathematical equivalence, which holds independently of physical considerations such as the dimension, the temperature, or even if the system is quantum or classical. However, this equivalence has a key physical consequence: it implies that $\alpha$ is observable in linear response measurements. In fact, high-frequency power spectra for quantum spin systems can be measured with nuclear magnetic resonance, and exponential decays were reported for CaF$_2$~\cite{PhysRev.188.609,PhysRevB.10.822,0953-8984-2-50-017}. This experimental technique therefore provides a practical way of measuring $\alpha$. On a theoretical note, the spectral function also appears in the off-diagonal Eigenstate Thermalization Hypothesis, which is therefore related to our hypothesis.
	
	Additionally, comparing \eqref{eq:spectral_fcn_bound} and \eqref{eq:spectral_density_asymptotic} shows that $\alpha \le \pi/2\kappa$, so the growth rate is limited by the local bandwidth of the model and the geometry:
	\begin{equation}
	    \alpha \le \pi e G_r ||h_i|| \,, \label{eq:alphabound}
	\end{equation}
   \textit{c.f.} \eqref{eq:spectral_fcn_bound}.	This inequality is the consequence of the natural energy scale for the Lanczos coefficients being set by the local bandwidth. However, we shall see that $\alpha$ itself is not merely the bandwidth, but contains a great deal of physical information about the system.

		We find it useful to dispel a possible misconception related to the high-frequency tail of the spectral function $\Phi(\omega)$. On dimensional grounds it is tempting --- though ultimately erroneous --- to interpret \eqref{eq:spectral_density_asymptotic} as a statement about the short-time behavior of $C(t)$. To see why this is wrong, notice that the short-time behavior is captured by the first moment alone, as $C(t) = 1 - \mu_2 \, t^2/2 + O(t^4)$. The high-frequency information instead governs the asymptotics of moments $\mu_{2n}$ as $n \to \infty$ {(which involve increasingly large operators)} and the analytical structure of $C(t)$ on the imaginary-$t$ axis, as shown in Fig. \ref{fig:strip}. In particular, the exponential decay rate sets the location of the closest pole to the origin on the imaginary axis. The high-frequency information also does not control the large time limit $t \to +\infty$; we will come back to this point in Section~\ref{sec:hydro_pheno} below. In brief, the hypothesis governs large $\omega$ behavior of $\Phi(\omega)$ and, correspondingly, the behavior of $C(t)$ on the \textit{imaginary} axis. Explicitly, a growth rate of $\alpha$ gives rise to a singularity at
		\begin{equation}
			t= \pm \frac{i \pi}{2\alpha} \,.
			\label{eq:autocorrelation_singularity}
		\end{equation}

	\subsection{Analytical Evidence}

    The upper bounds of the previous section show that the Lanczos coefficients cannot grow faster than linearly. We now show that this bound is tight through two analytic examples.

	It is an ironic point that the assumptions for the hypothesis \eqref{eq:linear_lanczos} fail in virtually all known solvable models, as those are often integrable, or even non-interacting. This explains why, to the best of our knowledge, linear growth was not recognized in any of the extensive literature on the recursion method as a universal behavior (except for certain classical systems~\cite{liu90XYZ}). However, there is one solvable model where we can compute the linear behavior analytically: the SYK model (see, e.g.~\cite{kitaev15,SY93,sykcomment}). Its Hamiltonian is
	\begin{equation}
	H_\text{SYK}^{(q)} = i^{q/2} \sum_{1 \leq i_1 < i_2 < \dots < i_q \leq N} J_{i_1\dots i_q}
	\gamma_{i_1} \gamma_{i_2} \cdots \gamma_{i_q}
	\label{eq:Q_SYK_model}
	\end{equation}
	where the $\gamma_i$'s, with $1\le i \le N$, are Majorana fermions with anti-commutators $\{\gamma_i, \gamma_j\} = \delta_{ij}$, and the $J_{i_1\dots i_q}$'s are disordered couplings drawn from a Gaussian distribution with mean zero and variance $(q-1)! J^2/N^{q-1}$. We study the dynamics of a single Majorana $\O = \sqrt{2} \gamma_1$~\cite{roberts2018operator}. To leverage the SYK solvability, we shall compute the moments $\mu_{2n} = \pbraket{\O \vert \mathcal{L}^{2n} \vert \O}$, averaged over disorder in the large-$N$ limit. For any finite $q$, the moments can be computed efficiently, thanks to the well-known large-$N$ Schwinger-Dyson type equations satisfied by the correlation functions.  The self-averaging properties of the SYK model allows the typical Lanczos coefficients to be computed from the averaged moments via a general numerical procedure~\cite{viswanath2008recursion}. This is described in detail in Appendix \ref{app:SYK}.
	
	We find that the Lanczos coefficients follow the universal form \eqref{eq:linear_lanczos} quite closely, as shown in Fig. \ref{fig:linear_lanczos_evidence}(a). In the large-$q$ limit, there is a closed form expression for the coefficients, computed in Appendix~\ref{app:SYK}:
	\begin{equation}
	b_n^{\text{SYK}} = \begin{cases} 
	\mathcal{J}\sqrt{2/q} + \BigO(1/q) & n = 1  \\
	\mathcal{J} \sqrt{n(n-1)} + \BigO(1/q) &  n > 1 \,,
	\end{cases}
	\label{eq:SYK_lanczos}
	\end{equation}
	where $\mathcal{J} =  \sqrt{q} \, 2^{(1-q)/2} J$. Therefore in the large-$q$ limit, the SYK model follows the universal form \eqref{eq:linear_lanczos} with $\alpha = \mathcal{J}$.  We may conclude that our hypothesis is obeyed in a canonical model of quantum chaos and that the upper bound of linear growth of the Lanczos coefficients is tight.
	%\Ehud{[Where do we define $\gamma$ and why is it needed here?]}\Xiangyu{eq 12; it's the offset.}

	The SYK model is quite unusual in several respects: it is a disordered, large-$N$ model in zero dimensions. However, none of these special features are required to achieve linear growth. To demonstrate this we turn to a model studied in the mathematical literature, defined on the 2d square lattice \cite{bouch2015complex}:
	\begin{equation}
		H = \sum_{x,y} X_{x,y} Z_{x+1,y} + Z_{x,y} X_{x,y+1}
		\label{eq:H_bouch}
	\end{equation}
	where $X$ and $Z$ are the normal Pauli matrices. A theorem \cite{bouch2015complex} states that the moments of the operator $X_{0,0}$ grow as
	\begin{equation}
		\mu_{2n} = n^{2n} e^{O(n)}
		\label{eq:moment_growth}
	\end{equation}
	which implies that the Lanczos coefficients grow linearly (see Appendix~\ref{app:recursion-review} for translation between asymptotics). Thus linear growth  \eqref{eq:linear_lanczos} is a tight-upper bound for the growth of the Lanczos coefficients in dimensions greater than one for ``realistic'' spin models. The content of our hypothesis is that achieving this upper bound is generic in chaotic systems.
%	\Thomas{[At this point of the paper, does the reader know that $\mu_{2n} \sim n^{2n} \leftrightarrow b_n \sim n$?]}\Xiangyu{yes; we quoted appendix A} 

	\subsection{The Special Case $d=1$}\label{sec:1dspecial}

\begin{figure}
		\centering
		\includegraphics{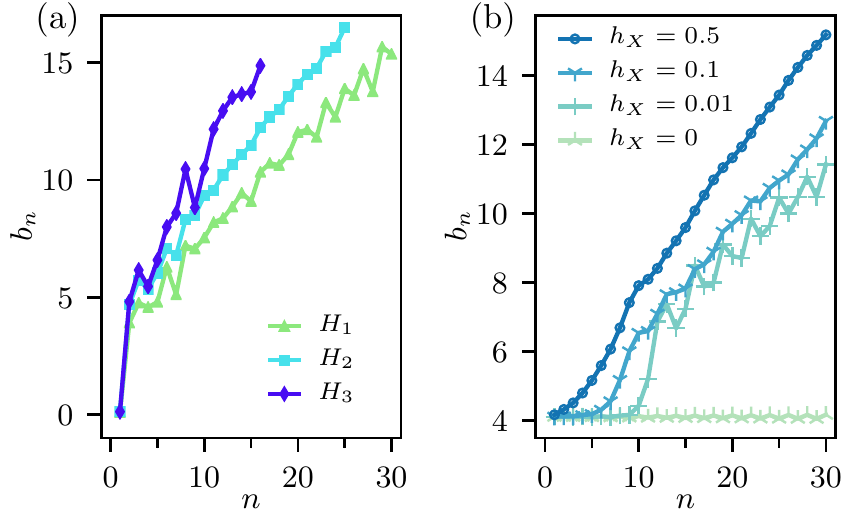}
		\caption{\textit{(a)} Lanczos coefficients in a variety of strongly interacting spin-half chains: $H_1 =\sum_i X_iX_{i+1} + 0.709 Z_i + 0.9045 X_i$, $H_2 = H_1 + \sum_i 0.2 Y_i$, $H_3 = H_1 + \sum_i 0.2 Z_i Z_{i+1}$. The initial operator $\O$ is energy density wave with momentum $q=0.1$. \textit{(b)} Cross-over to apparently linear growth as interactions are added to a free model. Here $H= \sum_{i} X_i X_{i+1} -1.05 Z_i + h_X X_i$, and $\O \propto \sum_i 1.05 X_i X_{i+1} + Z_i$. The $b_n$'s are bounded when $h_X = 0$ but appears to have asymptotically linear growth for any $h_X \neq 0$. Logarithmic corrections are not clearly visible in the numerical data. Numerical details are given in Appendix \ref{app:numerics}.}
		\label{fig:linear_lanczos_evidence}
	\end{figure}

We now turn to the special case of one dimensional systems.	Let us first present some numerical evidence. Fig. \ref{fig:linear_lanczos_evidence}(a) shows the Lanczos coefficients for a variety of spin models in the thermodynamic limit. (Numerical details are given in Appendix \ref{app:numerics}.) One can clearly see that the asymptotic behavior still \textit{appears} linear whenever the model is non-integrable. There is often an onset period before the universal behavior sets in; the first few Lanczos coefficients are highly model-dependent. We have observed that the more strongly-interacting the system, the sooner universal behavior appears~\footnote{This is quite fortuitous, computationally: as a general rule, in more strongly interacting systems, exponentially more parameters are required to compute a given $b_n$, so fewer $b_n$'s may be computed overall.}.  Fig.~\ref{fig:linear_lanczos_evidence}(b) shows the robustness of this asymptotic behavior. The pure transverse field Ising model may be mapped to free fermions so, as expected, the Lanczos coefficients are bounded. But as soon as a small integrability-breaking interaction is added, the coefficients appear to become asymptotically linear, and the asymptotic behavior sets in at smaller $n$ as the strength of the interaction increases. This is reminiscent of the crossover from Poisson to GOE distributed level statistics as integrability is broken~\cite{boyigas84,Ullmo2016}. Observe also that the slope of the asymptotic growth depends only weakly on the (integrability breaking) interaction strength. This seems to be a general phenomenon, as it occurs also in the SYK model plus two body interactions, see Fig.~\ref{fig:chaos_transition_SYK} for details.   %We have also checked a variety of other models believed to have chaotic behavior, and a number of operators in each. 

	%\Thomas{[old is subjective]}
	The numerical evidence is apparently compatible with linear growth of the Lanczos coefficients in 1d --- but only apparently. We can see this by considering the singularity structure of the correlation function. When the Lanczos coefficients achieve linear growth, there is a singularity in $C(t)$ on the imaginary axis, given by Eq. \eqref{eq:autocorrelation}. However, there is a classical theorem~\cite{araki1969gibbs} which says, roughly, that $C(t)$, $t \in \C$, is \textit{entire} for any local system in one dimension. Lanczos coefficients, therefore, must have strictly sublinear growth in one dimension. We note that this is an entirely geometric constraint, and has been previously noted by several works in a variety of contexts~\cite{abanin15,abanin17}, and derive it from first principles in Appendix \ref{app:geometric_bound_1d}.% \Thomas{[Need refs]}.

To formulate the hypothesis in one-dimension, we return to the informal version: \textit{the Lanczos coefficients should grow as fast as possible}. More concretely, the Lanczos coefficients should achieve the upper-bound imposed by the geometry. Following \cite{bouch2015complex}, we compute this bound in Appendix \ref{app:geometric_bound_1d} and can therefore formulate the hypothesis as follows. Suppose $H$ describes an infinite, non-integrable, many-body system in dimension $d$ and $\O$ is a local operator having zero overlap with any conserved quantity. Then the asymptotic behavior of the Lanczos coefficients is
\begin{equation}
	b_n =\begin{cases}
		A \frac{n}{W(n)} + \smallo(n/\ln n) \sim{} A \frac{n}{\ln n} + \smallo(n/\ln n) & d = 1\\
		\alpha n + \gamma + \smallo(1) & d > 1\\
	\end{cases}
	\label{eq:hypothesis_full}
\end{equation}
for some constants $\alpha, \gamma, A$ and $W$ is the Lambert $W$-function which is defined by the implicit equation $z=W(z e^{z})$ and has the asymptotic $W(n) = \ln n - \ln \ln n + o(1)$. In other words, the hypothesis acquires a logarithmic correction in one dimension. The coefficient $A$, like the growth rate $\alpha$, has dimensions of energy and can be bounded above by the local bandwidth; for Hamiltonians with nearest-neighbor local term $h_x$, we have (see Appendix~\ref{app:geometric_bound_1d})
\begin{equation}
    A \le \frac4e ||h_x|| \,. \label{eq:alphaboundwithlog}
\end{equation}
We note that, unlike in higher dimensions, we are not aware of any analytic examples which achieve the maximal growth rate in 1D, leaving open the possibility that the first line of \eqref{eq:hypothesis_full} is an over-estimate. 

In some sense, the linear growth ``barely breaks'' in one dimension; the Lanczos coefficients can still grow faster than $b_n \sim n^\delta$ for any $\delta < 1$. The phenomenological difference between linear growth in all dimensions and \eqref{eq:hypothesis_full} is often slight --- such as in Fig. \ref{fig:linear_lanczos_evidence}. Indeed, resolving logarithmic corrections in numerical data is a hard problem that often requires several decades of scaling. Altogether, we see that there is a subtle logarithmic correction to the operator growth hypothesis in one dimension.

	\section{Exponential Growth of Complexities}~\label{sec:complexity}
	
	Now that we have presented evidence in favor of the hypothesis, we shall turn to the analysis of its consequences. In this section we study the universal behavior of operators which have linear growth of Lanczos coefficients with rate $\alpha$. This is done in two steps. First, by studying the quantum mechanics problem \eqref{eq:1D_chain_problem} on the semi-infinite chain, we show that $\alpha$ measures the rate of exponential growth in operator complexity, in a sense we shall make precise below. Second, we prove that $\alpha$ gives an upper bound on a large class of {operator complexity measures}. Finally we shall remark on the case of linear growth with log-corrections.
	
	We remark that our notion of complexity is \textit{prima facie} distinct from other notions bearing the same name, such as circuit complexity (see the reviews \cite{aaronson2016complexity,susskindlecture} and references therein). Indeed, a satisfactory definition of operator complexity of any sort is an unresolved problem, and may not have a unique answer.
	
	\subsection{Exponential Growth in the Semi-infinite Chain} \label{sec:1dchain}
	Recall that the Lanczos algorithm reduces the operator dynamics to a discrete Schr\"odinger equation \eqref{eq:1D_chain_problem}, $$\partial_t \varphi_n = -b_{n+1} \varphi_{n+1} + b_n \varphi_{n-1}, \quad \varphi_n(0) = \delta_{n0} \,. $$ 
	We shall analyze this quantum mechanics problem when the hypothesis is satisfied in $d>1$, i.e. $b_n = \alpha n + \gamma + o(1)$.	
	
	\begin{figure}
		\centering
		\includegraphics{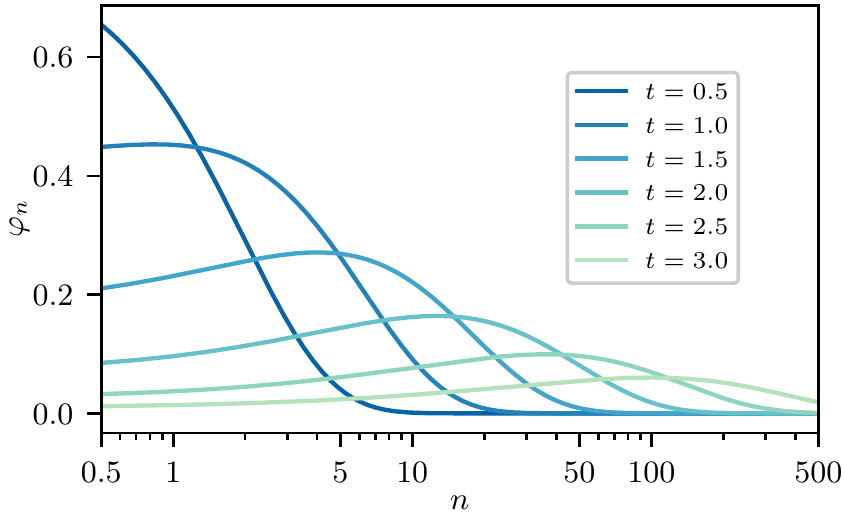}
		\caption{The exact solution wavefunction \eqref{eq:universal_wavefunction_main_text} in the semi-infinite chain at various times. The wavefunction is defined only at $n=0,1,2\dots$, but has been extrapolated to intermediate values for display.}
		\label{fig:exponential_spreading_wavefunction}
	\end{figure}
	
	As a first step, we take the continuum limit, by linearizing around momenta $0$ and $\pi$. This yields a Dirac equation $\partial_t \varphi = \pm 2 \alpha x \partial_x \varphi$, whose characteristic curves $x \propto e^{\pm 2 \alpha t}$ show the wavefunction spreads exponentially fast to the right in the semi-infinite chain with rate $2\alpha$. 
	We remark that among all power-law Lanzcos coefficient asymptotics $b_n \sim n^{\delta}$, the linear growth $\delta=1$ is the only one which results in exponential spreading. When $\delta  > 1$, the characteristic curves reach $x=\infty$ at finite time~\footnote{This seems non-physical and indeed, {has only been observed} in exotic classical systems~\cite{liu90XYZ}. It is ruled out whenever the dynamics are local by Eq. \eqref{eq:spectral_fcn_bound}.}. When $\delta < 1$, the spreading follows a power law $x \sim t^{1/(1-\delta)}$. In the case of $d=1$, with the logarithmic correction, the wavefunction spreads as a stretched exponential --- faster than any power law, but still slower than exponential. 
	%any supralinear growth of the Lanczos coefficients, such as $b_n \sim{} n^{1+\varepsilon}$, gives 
	
	To undertake a more careful analysis of the wavefunction on the semi-infinite chain, we employ a family of exact solutions. Suppose 
	\begin{equation}
	\widetilde{b}_n := \alpha \sqrt{n(n-1+\eta)} \; \xrightarrow{n\gg 1}\; \alpha n + \gamma,
	\label{eq:bn_exactly_solvable}
	\end{equation}
	where $\eta = 2\gamma/\alpha + 1$. For any system when the hypothesis is satisfied, the $b_n$'s will approach the $\widetilde{b}_n$'s asymptotically, so the properties of the exact solution using the $\widetilde{b}_n$'s are universal properties at large $n$. It is shown in Appendix~\ref{app:universal} that the full wavefunction for the operator evolving under the $\widetilde{b}_n$'s is
	\begin{equation}
	\pket{\O(t)} = \sum_{n=0}^\infty \sqrt{\frac{\left( \eta \right)_n}{n!}} \tanh(\alpha t)^n \sech(\alpha t)^\eta i^n\pket{\O_n}
	\label{eq:universal_wavefunction_main_text}
	\end{equation}
	where $\left( \eta \right)_n = \eta(\eta+1)\cdots (\eta+n-1)$ is the Pochhammer symbol and $\pket{\O_n}$ is the $n$th Krylov basis vector. Note that this example is not artificial but arises naturally in the SYK model, studied in Section~\ref{sec:sykchaos} below.
	
	The exact solution \eqref{eq:universal_wavefunction_main_text} benefits from a detailed analysis. Recall that the component of the wavefunction at some fixed site $n$ is $\varphi_n(t) = (-i)^n \pbraket{\O_n|\O_0(t)}$. For each $n$, $\varphi_n(t)$ is a purely real function which starts at $0$ (for $n>1$), increases monotonically until reaching a maximum at $t \sim{} \ln n$, then decreases as $\sim{}e^{-\alpha \eta t}$.  The fact that exponential decay, reminiscent of dissipative dynamics, emerges under unitary evolution is quite remarkable, and is only possible in an infinite chain~\footnote{This shows the importance of the thermodynamic limit. With any finite-dimensional Hilbert space, the chain would be finite, and the results in this section would be affected.}. Physically, the wavefunction is decaying by ``escaping'' off to $n\to\infty$, which serves as a bath. Note, however, that the hypothesis is not sufficient to show that $\varphi_n(t)$ decays exponentially with time for \textit{small} $n$, a fact whose consequences are studied in~\ref{sec:hydro_pheno} below. 
	
	We now come to a central consequence of the linear growth hypothesis: the exponential spreading of the wavefunction. At any fixed time and large $n$, the wavefunction $\eqref{eq:universal_wavefunction_main_text}$ has the form $\n{\varphi_n(t)}^2 \sim{} e^{-n/\xi(t)}$, where $\xi(t)$ is a ``delocalization length'' that grows exponentially in time: $\xi(t) \sim e^{2\alpha t}$ for $\alpha t\gg 1$. This exponential spreading is reflected in the expected position of the operator wavefunction \eqref{eq:universal_wavefunction_main_text} on the semi-infinite chain
	\begin{equation}
	\pbraket{n}_t := \pbraket{\O(t)|n|\O(t)} = \eta \sinh(\alpha t)^2 \sim{} e^{2 \alpha t} \,,
	\label{eq:expected_position}
	\end{equation}% \sout{cannot prove} [note: cannot implies that it is impossible, but we have merely been unable to prove it]
	More generally, $\pbraket{n^k}_t \sim e^{2 k \alpha t}$ for $k\ge 1$. This result agrees, of course, with the one obtained in the simple continuum-limit above. We believe that the asymptotic growth in \eqref{eq:expected_position} holds whenever the Lanczos coefficients grow linearly. Although we {have not proven} this assertion, we have checked that it holds for many cases, such as artificially generated sequences of Lanczos coefficients $b_n = \alpha n + \gamma_n$ with various kinds of bounded ``impurity'' terms $\gamma_n \sim \BigO(1)$. We will consider \eqref{eq:expected_position} as a fact that follows directly from the hypothesis: the position of an operator in the abstract Krylov space grows exponentially in time.
	
	We may interpret this exponential growth as a quantitative measure of the \textit{irreversible} tendency of quantum operators to run away towards higher ``complexity''~\cite{susskind2018things}. Indeed, we identify the position on the semi-infinite chain $\pbraket{n}_t$ as a notion of operator complexity. We refer to $\pbraket{n}_t$ as the ``Krylov-complexity" (or ``K-complexity'' for short) of an operator.
	After all, as $n$ increases, the operators $\O_n$ becomes more ``complex'', in the following sense: in the Heisenberg-picture, the equations of motions for $\O_n$'s form a hierarchy:
	\begin{equation}
	\begin{aligned}
	-i\dot{\O}_0(t) &= b_1 \O_1 (t) \,, \\
	-i\dot{\O}_1(t) &=  b_1 \O_0 (t) + b_2 \O_2(t)  \,,\\
	-i\dot{\O}_2(t) &=  b_2 \O_1 (t) + b_3 \O_3(t)   \,, \\[-0.1em]
	&\;\: \vdots
	\end{aligned}
	\end{equation}
	that is, the dynamics of $\O_n(t)$ depends on $\O_{n+1}(t)$.  This is analogous to the BBGKY hierarchy {in statistical mechanics}, in which the evolution of the $n$-particle distribution depends on the $(n+1)$-particle one. Similarly, as $n$ increases, the $\O_n$'s becomes less local in real space, involve more basis vectors in any local basis, and are more difficult to compute. 
	We remark that K-complexity is a distinct notion from precise terms such as circuit complexity and no relation should be inferred between the two. Closer precedents are the ideas of f-complexity and s-complexity~\cite{prosen11complexity}.
	
	We know from Section~\ref{sec:upper_bounds} that linearly growing Lanczos coefficients are the maximal rate so, in turn, the wavefunction may not spread faster than exponentially. Thus the hypothesis in $d>1$ implies that non-integrable systems have maximal growth of K-complexity: exponential, with rate $2\alpha$.

	\subsection{A Bound on Complexity Growth}
	\label{sec:complexity_bound}
	
	The physical meaning of K-complexity is far from transparent. After all, it depends on the rather abstract Krylov basis, the initial operator, and the choice of dynamics. To help pin down the idea of K-complexity, we study its relation to more familiar quantities. We shall consider a class of observables, ``q-complexities'' (q stands for \textit{quelconque}), that includes familiar notions like out-of-time-order correlators and operator size. We will show that the growth of any q-complexity is bounded above by K-complexity.
	
	We now define the q-complexity. Suppose $\mathcal{Q}$ is a superoperator that satisfies two properties:
	\begin{subequations}\label{eq:complexity_assumptions}
		\begin{enumerate}
			\item   $\mathcal{Q}$ is positive semidefinite. We denote its eigenbasis as $\pket{q_a}$, indexed by $a$, so that 
			\begin{equation}  
			\mathcal{Q} = \sum_a q_a \pket{q_a} \pbra{q_a}, \, q_a \ge 0 \,.  \label{eq:Q_basis}
			\end{equation}
			\item There is a constant $M > 0$ such that
			\begin{align}
			\pbraket{ q_b \vert \mathcal{L} \vert q_a}  &= 0 \text{ if } |q_a - q_b| > M \,, \label{eq:L_and_q} \\ 
			\pbraket{q_a \vert \O} &= 0 \text{ if } |q_a| >  M  \label{eq:O_and_q}\,.
			\end{align}
		\end{enumerate}
	\end{subequations}
	Then q-complexity is defined to be the expectation value 
	\begin{equation}
	\pbraket{\mathcal{Q}}_t := \pbraket{  \O(t) \vert \mathcal{Q} \vert  \O(t) },
	\end{equation}
	where $\O(t)$ is evolved under the Liouvillian dynamics of $\L$. A q-complexity is, in principle, an observable, and requires Hamiltonian (or Liouvillian) dynamics. {The rationale for the conditions is as follows: ~\eqref{eq:Q_basis} ensures the q-complexity is always non-negative,~\eqref{eq:L_and_q} guarantees it cannot change too much under one application of the Liouvillian, and~\eqref{eq:O_and_q} assigns a low complexity to the initial operator.} To illustrate this concept, we now consider three examples: K-complexity, operator size, and out-of-time-order correlators. \\[.2cm]
	\textbf{Example 1: K-complexity}.The K-complexity is always a q-complexity, with $\mathcal{Q} = \sum_n n \pket{\O_n} \pbra{\O_n}$. The basis $\pket{q_a}$ is just the Krylov basis $\pket{\O_n}$ and the conditions \eqref{eq:L_and_q} and \eqref{eq:O_and_q} are satisfied by construction of the Krylov basis with $M=1$. \\[.2cm]
	\textbf{Example 2: operator size}. A second example of a q-complexity is provided by \textit{operator size}~\cite{roberts2018operator}. For concreteness, we work in the framework of a spin-1/2 model (though Majorana fermions or higher spins work equally well). Consider the basis of \textit{Pauli strings}, e.g. strings $IXYZII\cdots$ with finitely many non-identity operators. Define $\mathcal{Q}$ to be diagonal in this basis, where the action of $\mathcal{Q}$ on a Pauli string is the number of non-identity Pauli's. So, for instance, $\mathcal{Q} \pket{IXYZI\cdots} = 3 \pket{IXYZI\cdots}$. The eigenvectors of $\mathcal{Q}$ have non-negative eigenvalues, so $\mathcal{Q}$ is positive semi-definite.
	
	Any choice of dynamics with at most $M$-body interactions (even long-ranged ones) will satisfy \eqref{eq:L_and_q}, while \eqref{eq:O_and_q} requires simple that $\O$ is $d$-local. So, under these conditions, the q-complexity $\pbraket{\mathcal{Q}}_t$ becomes the average size of Pauli strings contained in $\O(t)$:
	\begin{equation}
	\pbraket{\mathcal{Q}}_t = \sum_{\pi \in \text{Pauli strings}} \text{size}(\pi) \, \left|\pbraket{\pi \vert \O(t)}\right|^2 \,.
	\end{equation}\\[.2cm]
	\textbf{Example 3: OTOCs}. Our third --- and most interesting --- example of q-complexity is out-of-time-order commutators (OTOCs). Given $\O(t)$, each choice of local operator $V$ defines an OTOC $\pbraket{[V,\O(t)] \; | \; [V,\O(t)]}$. For simplicity, we work with a many-body lattice model, and consider an on-site operator $V_i$. We then define the OTOC superoperator by
	\begin{equation}
	\mathcal{Q} := \sum_i \mathcal{Q}_i, \quad \pbraket{A|\mathcal{Q}_i|B} := \pbraket{[V_i,A] \;\Big| \; [V_i,B]},
	\label{eq:OTOC_superoperator}
	\end{equation}
	where the sum runs over all lattice sites $i$. Provided the Hamiltonian and initial operator are $r$-local, and that the dimension $D$ of the on-site Hilbert space is finite, \eqref{eq:OTOC_superoperator} is a q-complexity. 
	
	To see this, let us work in the eigenbasis of $\mathcal{Q}$. For each site $i$, there is a basis $\mathcal{Q}_i \pket{q_{i,a}} = q_{i,a} \pket{q_{i,a}}$ with $1 \le a \le D^2$. We take $\pket{q_{i,0}}$ to be the identity operator with eigenvalue $0$, and note that $0 \le q_{i,a} \le Q$ for some finite $Q$. Since $[\mathcal{Q}_i,\mathcal{Q}_j] =\delta_{ij}$, the eigenbasis for the full operator space is the tensor product of the on-site bases. So for any sequence $\v{a} = \st{a_i}$, $\pket{q_{\v{a}}} := \otimes_i \pket{q_{i,a_i}}$ is an eigenvector satisfying
	\begin{equation}
	\mathcal{Q} \pket{q_{\v{a}}} = q_{\v{a}} \pket{q_{\v{a}}},\quad  q_{\v{a}} = \sum_i q_{i,a_i} \ge 0.
	\label{eq:OTOC_basis}
	\end{equation}
	For the eigenvalue to be finite, $a_i$ must be zero for all but a finite number of $i$'s and all eigenvalues are non-negative, so $\mathcal{Q}$ is positive semidefinite. Since the Hamiltonian is $r$-local, the matrix element $\pbraket{q_{\v{a}}|\L|q_{\v{b}}} \neq 0$ only if $\v{a}$ and $\v{b}$ differ on at most $r$ sites. So by \eqref{eq:OTOC_basis}, we may bound the difference $\n{q_{\v{a}} - q_{\v{b}}} \le M = r Q$. Similarly, any $r$-local operator satisfies \eqref{eq:O_and_q}. Having verified all the properties \eqref{eq:complexity_assumptions}, we may conclude that OTOCs of this form are a q-complexity. 
	
	OTOCs are known to be closely related to the operator size~\cite{roberts2018operator,maldacena2016bound}. It is usually possible to bound either quantity from the other, and to choose $V_i$ such that the OTOC reduces to the operator size.  
	
	We have now seen three examples of q-complexities, two of which are quantities that have been studied in recent times to understand the complexity of operators.  We remark that q-complexities (including K-complexity) are quadratic in $O(t)$ and \textit{not} linear response quantities, although the growth rate $\alpha$ is, via the spectral function. We will see in Section \ref{sec:classical} that q-complexities may also be applied to classical systems, though they work somewhat differently there. 

	A rigorous argument in Appendix \ref{app:bound} proves that, for any q-complexity,
	\begin{equation}
	\label{eq:Qtnt}
	\pbraket{\mathcal{Q}}_t \le  C \pbraket{n}_t,\, C = 2M \,.
	\end{equation}
	The following section will focus on the application of this general bound in the specific case of OTOCs. 

	To close this section, we show how the above results are affected by the log-correction to linear growth in 1d from Eq. \eqref{eq:hypothesis_full}: $b_n \sim A n / \ln n$. The continuum Dirac equation analysis yields a stretched exponential growth of K-complexity:
	\begin{equation}
	    (n)_t \sim e^{\sqrt{A t}} \,,
	\end{equation}
	which is slower than any exponential growth but faster than any power law. Combined with \eqref{eq:Qtnt}, we conclude that all q-complexities have at most stretched exponential growth in 1d.

	\section{Growth Rate as a Bound on Chaos}
	\label{sec:chaos_bound}
	
	We showed in the preceding section that K-complexity provides an upper bound for any q-complexity whatsoever, which includes certain types of OTOCs. Combining \eqref{eq:Qtnt} and \eqref{eq:expected_position}, we see that q-complexities grow at most exponentially in time, at least when the hypothesis holds for $d>1$. If that is the case, with $\pbraket{Q}_t \sim{} e^{\lambda_Q t}$, then the exponent is bounded above by $2\alpha$:
	\begin{equation}
	\lambda_Q \le 2 \alpha \,.
	\label{eq:bound_complexity}
	\end{equation}
	
	In the rest of this section we focus on the case where the q-complexity is an OTOC. When the OTOC grows exponentially at late times,
	\begin{equation}
	\pbraket{\mathcal{Q}^\text{OTOC}}_t \sim e^{\lambda_L t} \,, 
	\end{equation}
	its growth rate $\lambda_L$ is called the Lyapunov exponent, since in the classical limit it reduces to the Lyapunov exponent characterizing the butterfly effect in classical deterministic chaos \footnote{To be precise, the OTOC measures a \textit{generalized} Lyapunov exponent with $q=2$, which is greater or equal to the typical one~\cite{Politi13Lyapunov}}. We can then state following bound on Lyapunov exponents: \textit{for any system at infinite temperature where the operator growth hypothesis holds, then
	%	The bound \eqref{eq:Qtnt} then suggests the following {\textit{conjectural} bound on the Lyapunov exponent}:
	\begin{equation}
	\lambda_L  \le 2 \alpha, \label{eq:Lyapunov_conjecture}
	\end{equation}
	where we put $\lambda_L = 0$ whenever the OTOC grows slower than exponentially, and similarly for $\alpha$.} This follows directly from \eqref{eq:Qtnt} and \eqref{eq:expected_position}, so we have essentially proven \eqref{eq:Lyapunov_conjecture} as a mathematical proposition. 
	
	It is interesting to note that, as $\lambda_L$ is defined via a four-point correlation function (the OTOC), while $\alpha$ depends on a two-point correlation function ($C(t)$), the bound~\eqref{eq:Lyapunov_conjecture} can be interpreted as a relation between correlation functions of distinct nature. Such a relation is, to our knowledge, rather unusual (see \cite{murthy2019bounds} for a recent result). However, this point of view is not how we derived \eqref{eq:Lyapunov_conjecture}; an alternative proof working directly with the correlation functions would be illuminating. 
	
	Remarkably, the bound \eqref{eq:Lyapunov_conjecture} appears to be valid under much less restrictive assumptions --- at any temperature and in either classical or quantum systems. In this section, we examine the cases of quantum and classical systems at infinite temperature, and leave that of finite temperatures to Section \ref{sec:finite_temp} below.

	\subsection{SYK Model}\label{sec:sykchaos}

%	The Lyapunov exponents of SYK models are a topic of intense recent interest and have been computed in a variety of settings, including at infinite temperature in \cite{roberts2018operator}. In Appendix B we compute the Lanczos coefficients of the SYK model analytically in the large-$q$ limit and numerically for generic $q$. As expected, the Lanczos coefficients grow linearly for all $q$. Table \ref{tab:syk_exponents} shows that \eqref{eq:

	%At low temperatures $T = 1/\beta \ll J$, it is well-known that $\lambda_L = 2 \pi T$~\cite{kitaev15} saturates the universal quantum bound \eqref{eq:boundonchaos}. The finite-$T$ autocorrelation function \eqref{eq:autocorrelation_finite_T} with $\O = \sqrt{2} \gamma_1$ may be computed exactly by conformal invariance~\cite{sykcomment}:
	%\begin{equation} C(t) \propto \sech \left( t \pi T \right)^{2/q} \,. \end{equation}
%	This is the autocorrelation function of the exact solution \eqref{eq:universal_wavefunction_main_text}, and corresponds to Lanczos coefficients $b_n = \pi T \sqrt{n(n-1+\eta)}$ with $\eta = 2/q$ and $\alpha = \pi T$, in agreement with \eqref{eq:alpha_lowT}. Therefore the low-temperature SYK model saturates our bound~\eqref{eq:Lyapunov_conjecture}. 
	
%	At finite temperatures, using analytic results in the large-$q$ limit~\cite{sykcomment}, it is not hard to check (see Appendix~\ref{app:SYK}) that our bound is also saturated, whereas the universal bound~\eqref{eq:boundonchaos} is not.
	We illustrate the bound \eqref{eq:Lyapunov_conjecture} for the SYK model \eqref{eq:Q_SYK_model}. %, {where it holds at both infinite and low temperatures}.
	At infinite temperature, no analytic formula for the Lyapunov exponent is available, but it has been computed numerically in, e.g. \cite{roberts2018operator,sykcomment}. Table \ref{tab:syk_exponents} shows that not only does \eqref{eq:Lyapunov_conjecture} hold for the whole range of $q$-SYK models, but $\alpha$ is almost equal to $\lambda_L/2$, with exact agreement in the limit $q \to \infty$~\footnote{Indeed, the difference may well be a numerical effect, see \cite{roberts2018operator}.}.  These results show that the bound $\lambda_L \le 2\alpha $ is tight: the prefactor cannot be improved in general. Moreover, in the large $q$ limit, the probability distribution $\n{\varphi_n(t)}^2$ on the semi-infinite line is identical to the operator size distribution of $\gamma_1(t)$ \cite{roberts2018operator}. (See \eqref{eq:compare_syk} in Appendix \ref{app:SYK} for the precise statement.) So the large-$q$ SYK model is an instance where the quantum mechanics {problem} on the semi-infinite chain can be concretely interpreted. 
	
	We remark that in models with all-to-all interactions like SYK and its variants may be the only circumstances where the bound \eqref{eq:Lyapunov_conjecture} can be nearly saturated. For spatially extended quantum systems with finitely many local degrees of freedom, Lieb-Robinson bounds~\cite{Lieb2004} and its long-range generalizations~\cite{else2018improved} guarantee that the OTOC has slower-than-exponential growth in most physical systems at infinite temperature~\footnote{Indeed, generalized Lieb-Robinson bounds state that the OTOC between $\O(t)$ and $V_i$ is exponentially small if the site $i$ lies out of some volume which grows sub-exponentially. Then, a sum like~\eqref{eq:OTOC_superoperator} is essentially that volume.}.
	
	Such a difference can be understood as follows. Due to the lack of spatial structure in the SYK model, we expect operator complexity (by any reasonable definition) is almost completely captured by operator size which, in turn, is directly probed by OTOCs. In finite-dimensional systems, complexity should be a distinct concept from operator size. For instance, long Pauli strings generated in the non-interacting Ising models have nonetheless low complexity, since they can be transformed to simple few-body operators under the Jordan-Wigner transform. In non-integrable systems, by contrast, operator size growth is limited by Lieb-Robinson, while complexity can grow exponentially in the \textit{bulk} of an operator's support.

	\begin{table}[h]
		\centering
		\begin{tabular}{ccccccc}
			\toprule
			$q$ &2 & 3  & 4 & 7 & 10 & $\infty$ \\[0.2em] \hline
			$\alpha/\mathcal{J}$ & 0 & 0.461 & 0.623 & 0.800 & 0.863 & 1 \\  
			$ {\lambda_L}/ (2\mathcal{J})$ & 0 & 0.454  & 0.620 & 0.799 & 0.863 & 1 \\
			\botrule
		\end{tabular}
		\caption{The growth rate $\alpha$ versus half the OTOC-Lyapunov exponent $\lambda_L/2$ in the $q$-SYK model \eqref{eq:Q_SYK_model} in units of $\mathcal{J}=\sqrt{q}2^{(1-q)/2} J$. Here $\alpha$ is obtained by exact numerical methods discussed in Appendix~\ref{app:SYK}, while $\lambda_L$ is taken from the Appendix of~\cite{roberts2018operator}. The $q$-SYK is physical only for even integers $q$, but large-$N$ methods allow an extrapolation to any $q \geq 2$.}
		\label{tab:syk_exponents}
	\end{table}

	\subsection{Classical Chaos}\label{sec:classical}
	We now transition to the classical setting. After briefly explaining how the recursion method carries over almost verbatim to classical systems, we shall examine the classical form of the bound \eqref{eq:Lyapunov_conjecture}. However, the arguments of Section~\ref{sec:complexity_bound} do not carry over in full, and we are only able to prove a \textit{weaker} bound. We close with a numerical case-study that suggests the stronger conjectural bound may well be true (and tight).
	
	\subsubsection{A (Weaker) Bound on Classical Chaos}
	The recursion method applies to classical and quantum systems in exactly the same manner~\cite{viswanath2008recursion}. Classically, operator space is the space of functions on classical phase space and the Liouvillian $\mathscr{L} = i \{\mathscr{H},\cdot\}$ is defined by the Poisson bracket against the classical Hamiltonian $\mathscr{H}$ (we take $\hbar = 1$). The appropriate classical inner product at infinite temperature is $(f| g) = \int f^* g \, d\Omega$, where $d\Omega$ is the symplectic volume form on the phase space \footnote{We therefore require a compact phase space, such as in a classical spin model.}. The Liouvillian $\mathscr{L}$ is a self-adjoint operator, and the entire framework of the Lanczos coefficients carries over wholesale.
	
	Indeed, the Lanczos coefficients have been studied \textit{more} in the classical context. It is known~\cite{liu90XYZ,viswanath2008recursion} that linear growth of the Lanczos coefficients appears in general finite-dimensional, non-linear systems, to which we restrict ourselves~\footnote{Note that even if the phase space is finite-dimensional, the operator space is infinite-dimensional, allowing an infinite sequence of Lanzcos coefficients.}. The growth rate $\alpha$ is well-defined in such systems, as is the (classical) Lyapunov exponent $\lambda_L$, and the bound~\eqref{eq:Lyapunov_conjecture} takes on the same form as before: $\lambda_{L} \le 2\alpha$. In short, the similarity of classical and quantum Liouvillian evolution means that the recursion method --- and its consequences --- {carry over unchanged}.
	
	There is, however, one important caveat: a classical OTOC does \textit{not} generally qualify as a q-complexity. We will demonstrate this through an explicit, and instructive, example. Let us consider a single classical $SU(2)$ spin. Its classical phase space is the two-sphere, and classical operator space is spanned by the basis of spherical harmonics $\pket{Y^{m}_\ell}$, $\ell = 0,1,2\dots$, $m = -\ell, \dots, \ell$.
	
	A typical Hamiltonian is a polynomial of the classical spin operators $\mathscr{S}^x, \mathscr{S}^y, \mathscr{S}^z$ with Poisson bracket $\{\mathscr{S}^a, \mathscr{S}^b\} = - \varepsilon^{abc} \mathscr{S}^c$. We consider the simple non-linear example
	\begin{equation}
	\mathscr{H} = J \mathscr{S}^z \mathscr{S}^z + h_x \mathscr{S}^x.
	\label{eq:classical_example_Hamiltonian}
	\end{equation}
	Using Clebsch-Gordon coefficients one can show that the classical Liouvillian is quite sparse, and only the following matrix elements are non-zero:
	\begin{align}
	\pbraket{Y^{\ell\pm 1}_m|\mathscr{L}| Y^\ell_m} \neq 0 \,,\, 
	\pbraket{Y^\ell_{m \pm 1}|\mathscr{L}| Y^\ell_m} \neq 0,
	\label{eq:matrix_elements_cl}
	\end{align}
	whenever the states in question exist.
	
	We now examine the classical OTOC for the local operator $\mathscr{S}^z$,  given by matrix elements of a super-operator $\mathscr{Q}^z$. This operator is diagonal in the basis of spherical harmonics
	\begin{equation}
	\begin{aligned}
	\pbraket{ Y^k_n \vert \mathscr{Q}_z \vert  Y^\ell_m} 
	:=  & \pbraket{ \{\mathscr{S}^z, Y^{n}_k \}  \vert \{\mathscr{S}^z, Y^{m}_l \} } \\
	=&  \, m^2 \delta_{nm}\delta_{k\ell},
	\end{aligned}
	\label{eq:OTOC_classical}
	\end{equation}
	and we may immediately read off the eigenvalues as $m^2$. When $m$ changes by $1$ upon application of the Liouvillian, the eigenvalue $m^2$ changes by $1 \pm 2m$, which can be arbitrarily large. Hence the condition~\eqref{eq:L_and_q} cannot be satisfied for any finite constant $d$. It is {helpful} to recall that Section~\ref{sec:complexity_bound} showed the quantum OTOC is a q-complexity whenever the on-site Hilbert space is finite-dimensional. This fails in the case of a spin $s$, whose on-site dimension $2s+1$, in the classical limit $s\to \infty$. We have therefore seen that classical OTOCs are not q-complexities and, hence, the bound \eqref{eq:Lyapunov_conjecture} does not follow from the reasoning of Section~\ref{sec:complexity_bound} in the classical case, and remains a conjecture.
	
	Nonetheless, for any Hamiltonian and initial operators that are polynomials of the spin variables $\mathscr{S}^a$, we can show the following general bound
	\begin{equation}
	\lambda_L \leq 4 \alpha \,, \label{eq:weak_bound}
	\end{equation}
	which is weaker than the conjectured $ \lambda_L \leq 2 \alpha $.

	To show \eqref{eq:weak_bound}, observe that by \eqref{eq:OTOC_classical}, the superoperator $\mathcal{R}_z := \mathcal{Q}_z^\frac12$ satisfies \eqref{eq:L_and_q}, since its has eigenvalue $m$ for $Y^\ell_m$, which can change only by $\delta$ upon one Liouvillian application, where $\delta$ is the polynomial degree of the Hamiltonian. Other conditions in \eqref{eq:complexity_assumptions} are satisfied straightforwardly. We then have
	\begin{equation}
	e^{\lambda_L t} \sim  \pbraket{ \mathcal{Q}_z }_t = \pbraket{ \mathcal{R}_z^2 }_t \leq C^2 \pbraket{ n^2}_t  \sim  e^{4 \alpha t}
	\,,
	\end{equation}
	which implies \eqref{eq:weak_bound}. Here the first $\sim$ is by definition, the the inequality is a straightforward generalization of the bound on q-complexity, Eq.~\eqref{eq:Qktnkt} of Appendix~\ref{app:bound}, and the last $\sim$ is a generalization of \eqref{eq:expected_position} (see below that equation).
	
	This argument carries over to the OTOC with spin variables in any direction by spherical symmetry, and applies almost \textit{verbatim} to systems with a few spins, $\mathscr{S}_i^{x,y,z}, i = 1,\dots,N$. A Lyapunov exponent associated with a finite sum such as
	\begin{equation}
	\sum_{i=1}^N \sum_{a=x,y,z}  \pbraket{ \{\mathscr{S}_i^a, \O(t)\}  \Big| \{\mathscr{S}_i^a, \O(t) \} }
	\end{equation}
	satisfies the same bound since every term does so. In summary, \eqref{eq:weak_bound} is established in general classical few-spin models. We expect it is possible to show \eqref{eq:weak_bound} rigorously.
	
	An interesting corollary of \eqref{eq:weak_bound} is a relation between chaos and the decay rate of the spectral function. Recall that the linear growth of Lanczos coefficients is equivalent to the exponential decay of the spectral function $\Phi(\omega) \sim \exp(-|\omega|/\omega_0)$ at high frequency, where $\omega_0 = \tfrac{2}{\pi} \alpha$. Then \eqref{eq:weak_bound} is equivalent to
	\begin{equation} \label{eq:bound_spectral}
	\lambda_L \leq  2\pi \omega_0 \,.
	\end{equation}
	(The conjectured bound would {instead} imply $ \lambda_L \leq  \pi \omega_0$.) {In numerous classical systems, the power spectrum decay of time series has been used as an empirical probe of deterministic chaos  \cite{PhysRevA.23.2673,GREENSIDE1982322,segiti95chaos,sigeti95chaos1,cheskifov08lyapunov,elsayad14chaos,PhysRevLett.107.185003}. To the best of our knowledge, the bound \eqref{eq:bound_spectral} provides the first quantitative justification for this usage.}
	
	We mention that the relation between chaos and \textit{long-time} decay of correlation functions has also been studied: long-time relaxation to equilibrium was shown to be controlled by Ruelle resonances in specific chaotic models \cite{PhysRevLett.56.405,Turiaci2016}. However, the long-time and high-frequency behaviors are \textit{a priori} unrelated, as we discuss further in Section~\ref{sec:decay}. 
	
	We stress that the growth rate is an upper bound, but \textit{not} a diagnostic of \textit{classical} chaos. Indeed, our bound is correct but not tight for most classical integrable systems which, generically, have non-zero growth rate but no chaos ~\cite{liu90XYZ}. 
	
	Unfortunately, we are not able to improve the argument and prove the stronger conjectured bound. Instead, we resort to testing the validity of the conjectured bound~\eqref{eq:Lyapunov_conjecture} in a canonical example of classical chaos.

	\subsubsection{Numerical Case Study}
	The Feingold-Peres model of coupled tops~\cite{feingold1983regular} is a well-studied model of few-body chaos, both classically and at the quantum level~\cite{PhysRevA.30.509,fan2017quantum}.
	The quantum model is a system of two spin-$s$ particles, $1$ and $2$, with Hamiltonian
	\begin{equation}
	H_\text{FP} =  (1+c)\left[ S_1^z + S_2^z \right] + 4 s^{-1} (1-c) S_1^x S_2^x
	\label{eq:quantum_two_spin_model}
	\end{equation}
	where $c \in [-1,1]$ is a parameter and $S_i^{\alpha}$ satisfy the $SU(2)$ algebra $[S_i^\alpha, S_j^\beta] = i\hbar \delta_{ij} \varepsilon^{\alpha\beta\gamma} S_i^\gamma$ and act on a spin-$s$ Hilbert space. This is non-interacting when $c = \pm 1$ and chaotic in the intermediate region. Correspondingly, the Lanczos coefficients are asymptotic to a constant at $c = \pm 1$ and increase linearly in intermediate regions. However, since the operator space dimension is finite (equal to $(2s+1)^4$), the sequence of Lanczos coefficients is finite; in fact, the Lanczos coefficients saturate. The classical limit is obtained by taking $s$ to infinity. There the Hamiltonian becomes 
	\begin{equation}\label{eq:classical_two_spin_model}
	\mathscr{H}_\text{FP,cl} = (1+c)\left[ \mathscr{S}_1^z + \mathscr{S}_2^z \right] + 4 (1-c) \mathscr{S}_1^x \mathscr{S}_2^x 
	\end{equation}
	where $\mathscr{S}_i^\alpha, i = 1, 2$ are two sets of classical $SU(2)$ {spins}. As {an} $SU(2)$ representation, {the classical operator space} contains all integer spins, whereas the quantum operator space has only integer spins up to $2s$. 
	
\begin{figure}
		\centering
		\includegraphics{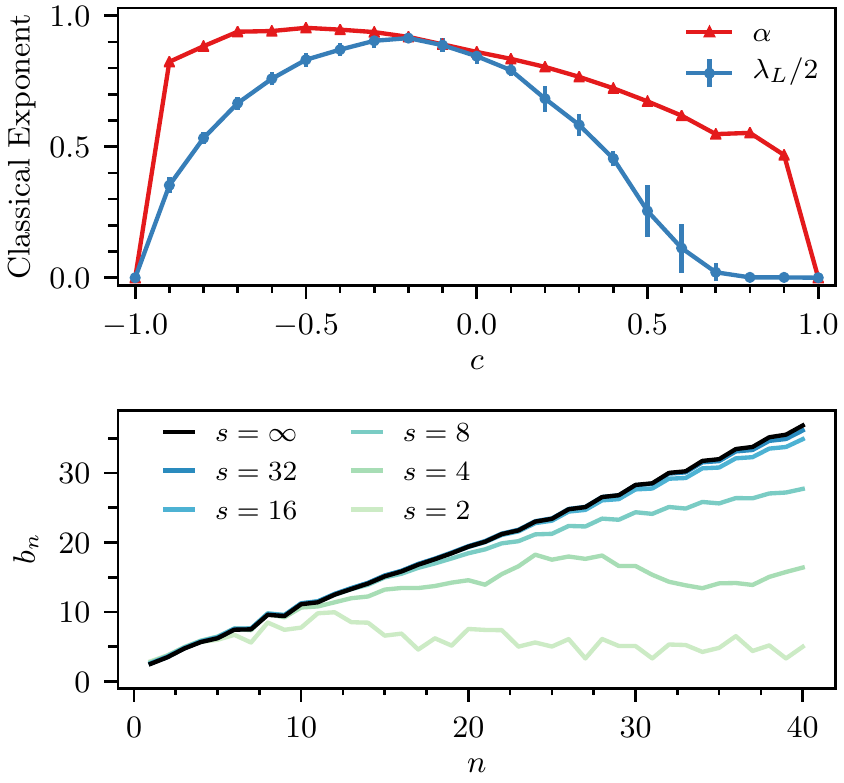}
		\caption{(a) The growth rate $\alpha$ versus the classical Lyapunov exponent $\lambda_L/2$ in the classical Feingold-Peres model of coupled tops, \eqref{eq:classical_two_spin_model}. $\alpha \geq \lambda_L/2$ in general, with equality around the $c=0$ where the model is the most chaotic. The growth rate appears to be discontinuous at the non-interacting points $c = \pm 1$, similarly to Fig.~\ref{fig:linear_lanczos_evidence}-(b).  (b) The first $40$ Lanczos coefficients of quantum $s = 2, \dots, 32$ and classical $(s = \infty)$ FP model, with $c = 0$. }
		\label{fig:classical_lyapunov}
	\end{figure}

	We compute the classical Lanczos coefficients for the operator $\O \propto S_1^z S_2^z$ ($ \mathscr{S}_1^z \mathscr{S}_2^z$ in the classical case). As shown in Fig.~\ref{fig:classical_lyapunov}(b), the quantum Lanczos coefficients converge to the classical ones as $s\to \infty$, as expected, and they increase linearly near $c=0$. We have checked that $\alpha$ does not depend on the choice of initial operator $\O$, so long as $\O$ does not overlap with any conserved quantity. 
	
	To test the conjectured bound \eqref{eq:Lyapunov_conjecture}, we compare the {growth rate} $\alpha$ with the classical Lyapunov exponent ($\lambda_L/2$ in our notation), which can be calculated by the standard variational equation method~\cite{alligood1996chaos}. Remarkably, the data shown in Fig. \ref{fig:classical_lyapunov}(a) corroborates the conjectured bound $\alpha \ge \lambda_L/2$ in the parameter region explored, with equality up to numerical accuracy in the regime $c\approx 0$, where the model is known to be maximally chaotic, with almost no regular orbits~\cite{PhysRevA.30.509,feingold1983regular}. Enlarging the parameter space, for instance by adding terms such as $\mathscr{S}^z_i$ to the Hamiltonian, give further results consistent with the bound. It is thus possible that the conjectured bound is valid in classical systems and becomes tight in highly chaotic ones.

	\section{Application to Hydrodynamics}\label{sec:decay}
	
	Structural information about quantum systems can enable numerical algorithms. As an example, the success of the density matrix renormalization group algorithm is a consequence of the area law of entanglement entropy~\cite{white1992,Osborne2002}. We now apply the hypothesis to develop a semi-analytical technique to calculate decay rates and autocorrelation functions of operators and, in particular, compute diffusion coefficients of conserved charges. The key idea is to use the hypothesis to make a meromorphic approximation to the Green's function. This section introduces the continued fraction expansion of the Green's function, describes the zoology of operator decay, and finally presents the semi-analytical method.

	\subsection{Continued Fraction Expansion: Brief Review}
	{We briefly review} the continued fraction expansion of the Green's function~\cite{viswanath2008recursion}. The Green's function \eqref{eq:Green_fcn_defn}
	%\begin{equation}
	%G(z) := \pbraket{\O \left\vert \frac{1}{z-\L} \right\vert \O} = 
	%\sum_{n=0}^{\infty} \frac{\mu_{2n}}{z^{2n+1}} \,,
	%\end{equation}
   is related to the autocorrelation $C(t)$ by the following transform:
	\begin{equation}
	G(z) = i \int_0^\infty C(t) e^{-izt} d t \,,\, C(t) = \oint G(z) e^{izt} \frac{d z}{2\pi i}  \,, \label{eq:laplace_transform}
	\end{equation}
	where the integration contour is taken to be the shifted real axis shifted down by $-i \epsilon$ for some small $\epsilon >0$. Since $C(t)$ is bounded on the real axis, $G(z)$ is analytic in the lower half-plane, but may contain singularities on the upper half plane. We shall refer to \eqref{eq:laplace_transform} as the Laplace transform, despite the fact that it differs from the usual definition by a  factor of $i$.
	
	In the Krylov basis, $G(z) = \left[ z-L \right]^{-1}_{00}$ corresponds to all paths that start on the first site, propagate through the chain, and return. We can divide all paths into those that stay on the first site, and those that first hop to the second site, propagate on sites $n \ge 2$, and then return. More formally, for each $n\ge0$, let $L^{(n)} := L_{p\ge n, q\ge n}$ be the hopping matrix on the semi-infinite chain restricted to sites $n$ and above, and let $G^{(n)}(z) := \left[ z-L^{(n)} \right]^{-1}_{nn}$ be the corresponding Green function. (Note that $G^{(0)}(z) = G(z)$.) {We then have the following} recursion relation --- hence the name ``recursion method'' --- 
	\begin{equation}
	G^{(n)}(z) = \frac{1}{z - b_{n+1}^2 G^{(n+1)}(z)} \,,\, n \geq 0 \,.
	\label{eq:recursive_green_fcn}
	\end{equation}
	(For a quick derivation~\cite{auerbach2018hall}, consider the polynomial $P_n(z) := \det(z - L^{(n)})$. By Cramer's rule we have $ G^{(n)}(z) = P_{n+1}(z) / P_n(z)$; a cofactor expansion gives 
	$P_{n}(z) = z P_{n+1}(z) - b_{n+1}^2 P_{n+2}(z)$. Then \eqref{eq:recursive_green_fcn} follows from the two preceding equations.)
	
	Applying Eq. \eqref{eq:recursive_green_fcn} recursively yields the continued fraction expansion:
	\begin{equation}
	G(z) = \dfrac{1}{z- \dfrac{b_1^2}{z-\dfrac{b_2^2}{z-\ddots}}}.
	\label{eq:green_fcn_recursive}
	\end{equation}
	To save space, we denote the recursion \ref{eq:recursive_green_fcn} by $G^{(n)} = M_{n+1} \circ G^{(n+1)}$, where $M_n$ is the M\"obius transform $w \mapsto 1/(z- b_n^2 w)$ and ``$\circ$'' denotes function composition. It is crucial that the convergence of the continued fraction expansions is quite subtle and quite different from the convergence of, say, Taylor series. Practically speaking, one can compute only a finite number of the $b_n$'s in most situations. Truncating the expansion by taking the rest of the $b_n$'s to be zero (or any constant) rarely provides a good approximation to the whole function \cite{viswanath2008recursion}.

	\subsection{Hydrodynamical Phenomenology}\label{sec:hydro_pheno}
	
	Long-time and large-wavelength properties of correlation functions are governed by emergent hydrodynamics. For each conserved charge (e.g. energy, spin), the density field should relax to equilibrium in a manner prescribed by a classical partial differential equation. Often this is a diffusion equation, though more exotic possibilities such as anomalous diffusion and ballistic transport (infinite conductivity) can also appear.
	
	A numerical (and sometimes experimental) protocol to probe the emergent hydrodynamics is to study the autocorrelation function of the density wave operator $\O_q = \sum_x e^{i q x} Q_x$ (here $Q_x$ is the operator of the conserved charge at $x$) at a range of momenta $q$. The behavior at large time is of especial interest, and can, in turn, be read off from the singularity structure of the Green's function. Let us give a few examples. If the closest pole to the origin is at $z = i\gamma$, then the autocorrelation function will decay exponentially as $e^{-\gamma t}$, while if the location of the closest pole varies quadratically as $z = i Dq^2/2$, then the dynamics are diffusive.  However, the presence of non-linear terms in addition to the linear diffusive ones can give rise to exotic behavior where the diffusion constant itself becomes a function of frequency. An example of this is $G(z) = \left[ z - i D(z) q^2 / 2 \right]^{-1}$, where {$D(z) = D_0 + D_1 \sqrt{z} $.} At any fixed $q$, $G(z)$ has a branch cut in addition the diffusive pole, so although the diffusion constant $D_0$ is still well-defined, autocorrelation functions decay \cite{Mukerjee2006} as a \textit{power law} in time~\footnote{We thank Achim Rosch for pointing out this possibility.}. Regardless, the full singularity structure of the Green's function determines the long-time behavior.

	Of course, {computing} the singularity structure of the Green's function is a {demanding} task. Even in integrable models, determining if the correct hydrodynamics is, say, diffusion or anomalous diffusion is non-trivial --- let alone computing diffusion coefficients (see Refs~\cite{medenjak2017lower, prosen17xxz, ilievski2018superdiffusion,gopalakrishnan2018kinetic,de2018diffusion,denardis18diffusion1} for recent developments). Indeed, accurately computing diffusion coefficients has been the goal of much recent numerical work \cite{ leviatan2017quantum, white2018quantum, hauschild2017finding}.  This difficulty is reflected in the continued fraction expansion \eqref{eq:green_fcn_recursive}: the location of the poles change with each new fraction, so the full analytic structure of $G(z)$ depends on \textit{all} of the $b_n$'s.
	
	Knowing that the coefficients obey the universal form $\eqref{eq:linear_lanczos}$ is not enough, because even though the wavefunction is spreading out into the semi-infinite chain exponentially fast, we are given no guarantee about the wavefunction at the origin $n=0$. For instance, the correlation functions $C_1(t) = \sech(\alpha t)$ and $C_2(t) = \left( 1 + t^2 \right)^{-\gamma}$ \cite{viswanath2008recursion} both correspond to Lanczos coefficients that grow linearly But $C_1(t)$ decays exponentially while $C_2(t)$ decays as a power law, so clearly the asymptotics of $b_n$ alone is insufficient to establish long-time behavior. The power law {decay} is nonetheless reflected in the Lanczos coefficients for $C_2(t)$, which have an alterating subleading tail. Precisely, they have the form $b_n = \alpha n + \gamma + (-1)^n f_n$ where the $f_n$'s are positive and decay to zero. Therefore determining the long-time tail of $C(t)$ probably requires additional information about the subleading corrections to the hypothesis. In particular, the results in this work are \textit{prima facie} unrelated to a bound on transport~\cite{hartnoll}.

	%~\footnote{This presupposes, in particular, that autocorrelation functions decay exponentially in time.}
	\subsection{Numerical Diffusion Coefficients}
	
	Despite the complex behavior of autocorrelation functions in the time domain, there are situations where the hypothesis alone suffices to compute diffusion coefficients. In the case where the $b_n$'s approach the universal form \eqref{eq:linear_lanczos} especially quickly and regularly, we are able to make a meromorphic approximation to $G(z)$. The idea is as follows. In the semi-infinite chain picture, we may hope to calculate the first few Lanczos coefficients exactly, so we may describe behavior near the origin $n=0$ exactly. For large $n$, on the other hand, the hypothesis gives the coefficients almost exactly, so we can describes the dynamics by some exact solution. By stitching the dynamics at large and small $n$ together, we can hope to find the dynamics on the whole chain. This allows us to recover a diffusive dispersion relation and numerically extract the diffusion constant in specific models. 
	
	We remark that there are a number of existent extrapolation schemes to determine the Green's function from the first few Lanczos coefficients~\cite{viswanath2008recursion,auerbach2018hall}. The new ingredient here is the hypothesis, which controls the approximation.

	To make this idea into a precise numerical technique, we need three ingredients: a way to compute the Lanczos coefficients at small $n$, an exact solution at large $n$, and a robust way to meld them together. For a 1D spin chain in the thermodynamic limit of large system size, it is straightforward to compute the first few dozen Lanczos coefficients exactly through repeated matrix multiplication. Details are given in Appendix \ref{app:numerics}.
	
	To find the large $n$-behavior, we employ an exact solution for the quantum mechanics problem on the semi-infinite chain. If the hypothesis is obeyed, then the $b_n$'s also asymptotically approach the form
	\begin{equation}
	\widetilde{b}_n = \alpha \sqrt{n(n-1+\eta)} \xrightarrow{n\gg 1} \alpha n + \gamma, 
	\label{eq:universal_bn}
	\end{equation}
	where $\eta = 2\gamma/\alpha + 1$. The agreement is better, of course, at large $n$. The coefficients $\widetilde{b}_n$ have the virtue that the quantum mechanics problem they describe on the semi-infinite chain is exactly solvable. Appendix \ref{app:universal} applies the theory of Meixner orthogonal polynomials of the second kind to determine the autocorrelation analytically: $C(t) = \sech(\alpha t)^\eta$. (This is the same exact solution used in Section \ref{sec:complexity} above.) By Laplace transform, the corresponding Green's function is
	\begin{subequations} 
		\begin{align}
		\widetilde{G}_{\alpha,\gamma}(i z) &= \frac{1}{\alpha} H(z/\alpha; \eta),\\
		H(z; \eta) &= \frac{2^\eta}{z+\eta} \pFq{1}{2}(\eta, \frac{z+\eta}{2}, \frac{z+\eta}{2}+1; -1),\\
		\widetilde{G}^{(n)}(z) \ &=\  \widetilde{M_n}^{-1} \circ  \cdots \circ \widetilde{M_1}^{-1} \circ \widetilde{G}(z)
		\end{align}
		\label{eq:universal_green_fcn}
	\end{subequations}
	Here $\pFq{1}{2}$ is the hypergeometric function and $\widetilde{M_k}$ depends on $\widetilde{b}_{k}$.  It is crucial that $\widetilde{G}^{(n)}(z)$ is known analytically, so that \eqref{eq:universal_green_fcn} provides the asymptotically exact large $n$-behavior.
	
	Now we stitch the small and large $n$ information together. The true Green's function $G^{(N)}(z)$ only depends on the coefficients $b_n$ with $n \ge N$. So for sufficiently large $N$, where the $b_n$'s are approximately the same as the $\widetilde{b}_n$'s, we may approximate
	\begin{equation}
	\begin{aligned}
	G(z) &= M_1 \circ \cdots \circ M_{N} \circ G^{(N)}(z)\\
	& \approx M_1 \circ \cdots \circ M_{N} \circ \widetilde{G}^{(N)}_{\alpha,\gamma}(z),
	\end{aligned}
	\label{eq:G_universal_approximation}
	\end{equation}
	an approximation that becomes better at large $N$. Equation \eqref{eq:G_universal_approximation} is our semi-analytical approximation to the Green's function. One can check that this is a meromorphic approximation for $G(z)$, whose poles lie only in the upper half plane.

\begin{figure}[h]
		\centering
		\includegraphics{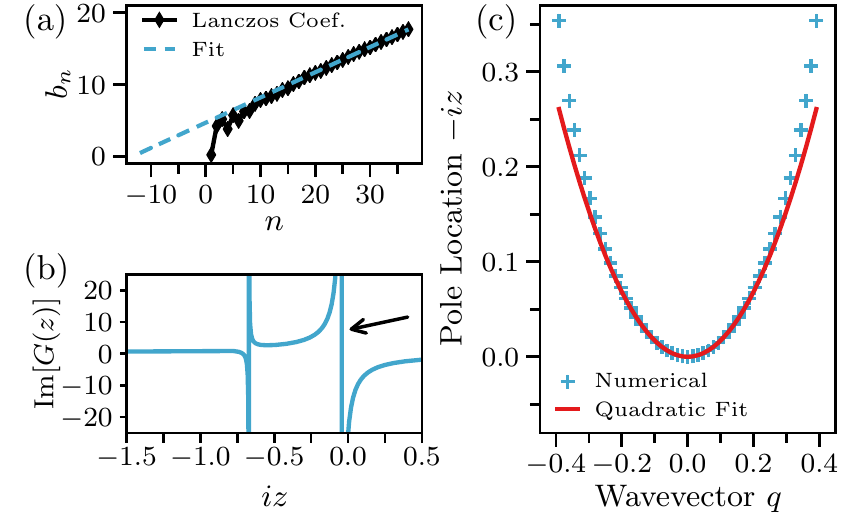}
		\caption{
			Numerical computation of the diffusion coefficient for the energy density operator $\O = \mathcal{E}_q$ in $H = \sum_{i} X_i X_{i+1} - 1.05 Z_i + 0.5 X_i$. {(a)} The Lanczos coefficients for $q = 0.15$ are fit to \eqref{eq:universal_bn} with $\alpha = 0.35$ and $\eta = 1.74$. We found it actually better not to approximate $G^{(N)}(z)$ by $\widetilde{G}^{(N)}(z)$, but instead by $\widetilde{G}^{(N+\delta)}(z)$ for some integer offset $\delta$ so that $\eta \approx 1$ (in the example shown, $\delta = 12$). Large $\eta$ or negative values lead to numerical pathologies.  {(b)} The approximate Green's function \eqref{eq:G_universal_approximation} at $q = 0.15$. The arrow shows the ``leading'' pole that governs diffusion. {(c)} The locations of the leading poles for a range of $q$. One can clearly see the diffusive dispersion relation $z = i Dq^2 / 2 + \BigO(q^4)$. Fitting yields a diffusion coefficient $D =3.3(5)$. 	
		}
		\label{fig:numerical_diffusion}
	\end{figure}

	In practice, one must calculate the $b_n$'s until the universal behavior appears and fit $\alpha$ and $\eta$. Then the approximate $G(z)$ can be calculated from \eqref{eq:universal_green_fcn} and a sequence of two-by-two matrix multiplications. One can then find the location of the first pole on the imaginary axis for a range of wavevectors $q$ and fit $z = iDq^2 / 2 + O(q^4)$ to extract the diffusion coefficient $D$. This procedure is illustrated for the energy diffusion in chaotic Ising model in Fig. \ref{fig:numerical_diffusion}. Almost all the computational effort goes into in computing the first few $b_n$'s exactly. We also note that the extrapolation is carried out with a linear fit to the Lanczos coefficients which is not strictly appropriate to $d=1$ (the log-correction is missing). Nevertheless, the numerical value of the diffusion coefficient appears to match other methods to within a few percent.\footnote{We are greatful to Francisco Machado and Biantian Ye for sharing their density matrix truncation (DMT) results with us.}	Further numerical tests on this example indicate the the exact asymptotics of Lanczos coefficients may not be necessary to compute $D$ to a decent precision.

	In short, the hypothesis is sometimes sufficient to describe the emergent hydrodynamic behavior of operators, even if we ignore the log correction in 1d. We reiterate that the hypothesis governs the leading order asymptotics of the Lanzcos coefficients only, while the autocorrelation depends on {further corrections}, so there is no \textit{a priori} reason it should be computable just from the hypothesis. On the other hand, in the better scenarios, \textit{less} knowledge on the Lanczos coefficients is required to capture the hydrodynamic coefficients. We will provide further examples of this algorithm and discuss its theoretical and practical accuracy in subsequent work.

	\section{Finite Temperature}
	\label{sec:finite_temp}

%	So far our discussion has been confined to infinite temperature. This section discusses the generalization to finite temperature and some associated results. Throughout, we denote the temperature by $T = \beta^{-1}$ and set $k_\mathrm{B} = 1$.

So far our discussion has been confined to infinite temperature. In this section we generalize to finite temperature. Only a minor modification is required to carry out the Lanczos algorithm at finite temperature so many of our results carry over unaffected. A summary is provided in Table \ref{tb:finite_infinite_temp_correspondence} for the reader's convenience. %  --- but not all

\begin{table}[h]
    \centering
    \begin{ruledtabular}
    \begin{tabular}{lc ll}
&  & $T=\infty$ & $T < \infty$\\[0.3em] \colrule
Inner Product & & $\pbraket{A|B}  \propto \Tr[A^\dagger B]$ & Eq. \eqref{eq:inner_products_finite_temp}\\
Lanczos Algorithm & & Eq. \eqref{eq:Lanczos_algorithm} & Eq. \eqref{eq:Lanczos_algorithm_finite_T}\\
$C(t), G(z), \Phi(\omega), \mu_{2n}$ & & Section \ref{sec:review} & Eq. \eqref{eq:dynamical_quantities_finite_T}\\
$b_n \leftrightarrow  C \leftrightarrow G \leftrightarrow \Phi \leftrightarrow \mu$  & & App. \ref{app:recursion-review} & App. \ref{app:recursion-review}\\
Hypothesis & & Eq. \eqref{eq:hypothesis_full} & Eq. \eqref{eq:hypothesis_finite_temp}\\
$b_n \sim{} \alpha n$ for SYK && Eq. \eqref{eq:SYK_Lanczos_coeffs} & Eq. \eqref{eq:alpha_SYK_q}\\
%K-complexity Defn. & ($\star$) & Eq. \eqref{eq:expected_position} &  Eq. \eqref{eq:expected_position}\\
%Q-complexity Defn. &(X) & Eq. \eqref{eq:complexity_assumptions} & Unspecified\\
Bound $\lambda_L \le 2 \alpha$ && Proven & Conjectured\\
\end{tabular}
    \end{ruledtabular}
    \caption{Correspondence between finite and infinite temperature definitions and results.  }%Here ($\star$) indicates a direct generalization where only the inner product is modified. Conversely, (X) indicates a significant difference. }
    \label{tb:finite_infinite_temp_correspondence}
\end{table}

    \subsection{Choice of Inner Product}
%	 Once this choice has been made, the Lanczos algorithm \eqref{eq:Lanczos_algorithm} can be run as usual, generating the finite-temperature Lanczos coefficients. We will see that though the \textit{values} of the Lanczos coefficients are modified by the finite-temperature inner product, the \textit{relationships} between the Lanczos coefficients and the spectral function are independent of the choice of inner product and are therefore unchanged. 

A single modification is required to adapt the formalism of recursion method to finite temperature: an operator inner product which incorporates the thermal density matrix. At temperature $T = 1/\beta$ (we set $k_{\text{B}}=1$), a general operator scalar product is defined by the
integral~\cite{viswanath2008recursion}:
	\begin{equation}
	\begin{aligned}
	\pbraket{A|B}_{\beta}^{g} &:= \frac1Z\!\int_{0}^\beta  g(\lambda) 
	\Tr[y^{\beta-\lambda} A^\dagger y^{\lambda} B] \, d\lambda
%	&\, =\int_0^\beta g(\lambda)\langle A^\dagger(i\lambda) B\rangle_\beta \,d\lambda \,,
	\end{aligned}
	\label{eq:inner_products_finite_temp}
	\end{equation}
	where $g(\lambda)$ is some even function on the thermal circle $[0,\beta]$, $y := e^{-H}$, and $Z := \Tr[y^\beta]$~\footnote{Precisely, $g$ must satisfy $g(\lambda) \ge 0$, $g(\beta -\lambda) = g(\lambda)$, and $\beta^{-1} \int_0^\beta d\lambda g(\lambda) = 1$. We also restrict to the subspace of operators with zero thermal expectation value, and omit the disconnected term in \eqref{eq:inner_products_finite_temp}.}. The choice of the inner product is not arbitrary, but is equivalent to the choice of the correlation function
	\begin{equation}
	    C_\beta^{g}(t) = \pbraket{\O|\O(t)}_\beta^{g} = 
	     \int_{0}^\beta  g(\lambda) 
	\mathrm{Tr}[ \rho_\beta \O^\dagger  \O(t+ i \lambda)] d\lambda \label{eq:Ct_finiteT}
	\end{equation}
	(where $\rho_\beta = e^{-\beta H} / Z$), which is in turn determined by the physical context; in fact, only a few choices of $g$ are physically relevant, such as \eqref{eq:inner_finite_T_CMT} and \eqref{eq:inner_finite_T} below.
	% through the choice of the function $g(\lambda)$, affects the Lanczos coefficients, the spectral function, and all the other quantities enumerated in \eqref{eq:equiv_dynamics}; yet, the relations\textit{ between} them are independent of the inner product. As $T \to \infty$, any choice of $g$ reduces to the inner product $\pbraket{A|B} = \Tr[A^\dagger B]/\Tr[1]$ standard in the previous sections. %At finite $T$, the different choices are mathematically related in a well-known way (see below). 

Once the inner product is chosen, the Lanczos coefficients are defined by the same Lanczos algorithm with the new norm. Quite explicitly, the recursion is:
	\begin{equation}
	\begin{aligned}
	\pket{A_n} &:=\L  \pket{ \O_{n-1}}_{\beta}^{g}- b_{n-1,T}^{(g)} \pket{ \O_{n-2}}_{\beta}^{g} \,, \\
     b_{n,T}^{(g)} &:= [\pbraket{A_n \vert A_n}_{\beta}^{g}]^{1/2} \,, \\
	\pket{\O_{n}}_{\beta}^{g} \ &:= \ \left( b_{n,T}^{(g)} \right)^{-1} \; \pket{A_n} \,, \\
	\end{aligned}
	\label{eq:Lanczos_algorithm_finite_T}
	\end{equation}
for $n=1, 2, 3, \dots,$ starting from $\pket{\O_{0}}_{\beta}^{g} := \pket{\O}$, $\pket{\O_{-1}}_{\beta}^{g} := 0$ and $b_{0,T}^{(g)} := 0$. We emphasize that \textit{only} the inner product has been changed compared to the infinite-$T$ version. In fact, the Krylov subspaces $\mathrm{span}\{\pket{\O}, \L \pket{\O} , \dots, \L^n \pket{\O}\}$ are unchanged at finite temperature, and only the notion of orthogonality is different, giving us a new orthogonal basis for those spaces. Also, we have the same \textit{relationships} between the Lanczos coefficients and the correlation function~\eqref{eq:Ct_finiteT}, as well as its linear transforms, the Green's function and spectral function 
	\begin{subequations}
	\begin{align}
	     G_{\beta}^{g}(z) &:= i \int_0^{\infty} e^{-izt} C_\beta^{g}(t) dt, \label{eq:G_finiteT}\\
	    \Phi_\beta^{g}(\omega) &:=  \int_{-\infty}^{\infty} e^{-i\omega t} C_\beta^{g}(t) dt \, , 
	   % \mu_{2n,T}^{(g)} &:= \pbraket{\O|\L^{2n}|\O}_{\beta}^{g}  \,.
	\end{align}
	    \label{eq:dynamical_quantities_finite_T}
	\end{subequations}
where the superscript $g$ is not an exponent. For example, the Green function \eqref{eq:G_finiteT} admits the continuous fraction expansion
\begin{equation}
	G(z) = \dfrac{1}{z- \dfrac{ \Delta_{1,T}^{(g)}   }{z-\dfrac{ \Delta_{2,T}^{(g)}   }{z-\ddots}}} \,,\, \Delta_{n,T}^{(g)} := \left(b_{n,T}^{(g)} \right)^2 \,,   
	\label{eq:green_fcn_recursive_finiteT}
	\end{equation}
which is identical to \eqref{eq:green_fcn_recursive}, except that $b_n$ are replaced by the finite-$T$ Lanczos coefficients. Similarly, the results of Appendix \ref{app:recursion-review} carry over directly.  %The general rule of thumb is that we can promote $b_n \to b_{n,T}^{(g)}$ and all our previous results carry over directly. The exceptions to this rule are given in Table \ref{tb:finite_infinite_temp_correspondence}.
	
The statement of the hypothesis at finite temperature is also directly analogous. We hypothesize that a chaotic system should have maximal growth of the Lanczos coefficients,
\begin{equation}
	    b_{n,T}^{(g)} = \alpha_T^{(g)} n + \gamma + o(1),
	   \label{eq:hypothesis_finite_temp}
\end{equation}
under the same conditions as before. Here $\alpha_T^{(g)} \ge 0$ depends on the inner product. Evidence for the hypothesis at finite $T$ will be provided in Section~\ref{sec:SYK_finiteT}.

Though the Lanczos algorithm proceeds in the same way for any choice of inner product, this choice will determine what physical correlation function we end up computing. There are two prominent choices of inner products: 
  \begin{itemize}
      \item In linear response theory, we use the ``standard'' inner product given by $g(\lambda) =  [\delta(\lambda) + \delta(\lambda - \beta) ]/2$:
    \begin{equation}
	(A | B)_\beta^{S} :=  \frac1{2Z} \Tr[y^\beta A^\dagger  B +    
	 A^\dagger y^\beta  B ] \,
	\label{eq:inner_finite_T_CMT} 
    \end{equation}
    that leads to the usual thermal correlation function.
  \item In quantum field theory, it is often natural to consider the Wightman inner product, which corresponds to $g(\lambda) = \delta(\lambda-\beta/2)$:
	\begin{equation}
	(A | B)_\beta^{W} :=  \frac1Z \Tr[y^{\beta/2}  A^\dagger y^{\beta/2} B ] \,. \label{eq:inner_finite_T} 
	\end{equation}
	In particular, this inner product allows us to relate our bound on chaos~\eqref{eq:Lyapunov_conjecture} and the finite-temperature bound of Ref.~\cite{maldacena2016bound}.
  \end{itemize} 
In equations~\eqref{eq:inner_finite_T_CMT} and \eqref{eq:inner_finite_T} and below, we replace the $g$ by $S$ or $W$ to indicate the choice of standard and Wightman inner product, respectively. At infinite temperature, both inner products reduce to the one $\pbraket{A|B} = \Tr[A^\dagger B]/\Tr[1]$ considered previously.
    
 The spectral functions of the two choices are related by a well-known identity: 
\begin{align}
	\Phi_\beta^{W} (\omega)= &\sech\left(\frac{\omega\beta}{2}\right) 	\Phi_\beta^{S}(\omega)  \xrightarrow{\omega \gg T}  e^{-\beta\omega/2} 	\Phi_\beta^{S}(\omega) \,,
	\label{eq:standard_vs_Wightman}
\end{align}
which follows directly from the definition \eqref{eq:spectral_fcn_defn}.  The Wightman inner product therefore imposes an extra temperature-dependent exponential decay to the spectral function, due to the suppression of high energy excitation by the two $e^{-\beta H/2}$ factors in \eqref{eq:inner_finite_T}. This observation will be crucial in the following section. On the other hand, it would be very interesting to understand how the high-frequency tail of $\Phi(\omega)_{\beta}^S$ depends on the temperature.

%We have seen that the generalization of our results is plagued by a mathematical ambiguity in the choice of inner product. One possible choice links our results to the bound on chaos of  ~\cite{maldacena2016bound}, and seen that our conjecture \eqref{eq:Lyapunov_conjecture} appears to generalize it to all temperatures. The choice of inner product is a physical choice and the best choice likely depends on the situation at hand.  %This well understood mathematical ambiguity must be fixed by physical considerations.  Nevertheless, different choices are related 

   \subsection{Bound on Chaos}
%   \Dan{I've re-written this making temperature an explicit variable in all symbols for clarity.}
A key result on quantum chaos at finite temperature is the bound on chaos of Ref. ~\cite{maldacena2016bound}. This universal bound was derived for quantum field theories at finite temperature $T = \beta^{-1}$, and reads as follows
\begin{equation}
\lambda_{L,T} \le 2\pi T \,\label{eq:boundonchaos}\end{equation} 
in natural units $\hbar = k_\mathrm{B}= 1$. It is nontrivial in finite-temperature quantum systems, and is therefore complemented by our bound {$\lambda_{L} \le 2 \alpha$} \eqref{eq:Lyapunov_conjecture} which applies to infinite temperature quantum and classical system. This leads to two natural questions. Can our bound be extended to finite temperature? How does it compare to the universal one?

Since {$\alpha_T^{(g)}$} depends on the inner product, and the finite-$T$ OTOC admits various regularizations, it is already a nontrivial task to find the correct formulation of the extension. To make progress  we consider the regularization scheme used for four-point OTOCs in~\cite{maldacena2016bound} to derive the universal bound. This scheme inserts the operators in the thermal circle $[0,\beta)$ with even spacing, as does the Wightman inner product \eqref{eq:inner_finite_T}. This suggests that an extension of the bound $\lambda_L \le 2 \alpha$ to finite temperature can be obtained by comparing the finite-T Lyapunov exponent (as defined in~\cite{maldacena2016bound}) and the finite-T growth rate defined with the Wightman inner product:
\begin{equation}
  \lambda_{L,T} \le 2 \alpha_{T}^{(W)} \quad \text{(conjecture)}
    \label{eq:conjecured_bound} \,.
\end{equation}
We stress that this is a conjecture below infinite temperature. Nevertheless, as we show in Section~\ref{sec:SYK_finiteT} below, exact results in the $q$-SYK model suggest that \eqref{eq:conjecured_bound} is plausible and tight. % The proof in Section~\ref{sec:complexity_bound} does not carry over  due to the presence of thermal inner product, the eigen-operators of the OTOC superoperator $\mathcal{Q}$ may not be operators with strictly finite support, but may contain decaying tails of arbitrarily long strings. Thus, the conditions \eqref{eq:L_and_q} may be too rigid and not satisfied, and one would need to relax them to account for the tails. Nevertheless, 
	
We now turn to the relation between the conjecture~\eqref{eq:conjecured_bound} and the universal bound, and show that the former infers the latter. By \eqref{eq:standard_vs_Wightman}, the Wightman spectral function decays at least as fast as $e^{-\beta\omega/2}$ at high frequency (because $\Phi_{\beta}^S(\omega) \le 1$). By \eqref{eq:spectral_density_asymptotic}, this is equivalent to the following upper bound on the Lanczos coefficients growth rate: 
\begin{equation}
    \alpha_T^{(W)} \le \pi T \label{eq:alpha_lowT} \,,
\end{equation}
where $ \alpha_T^{(W)} $ denotes the growth rate with Wightman inner product. Therefore, the conjecture {\eqref{eq:conjecured_bound}},
if true, would be tighter than the universal one {$\lambda_{L,T} \le 2\pi T$} \eqref{eq:boundonchaos}. At low temperature ($\beta \to \infty$ limit), the decay of $ \Phi_{\beta}^{W}(\omega)$ is dominated by the factor $e^{-\beta\omega/2}$, so {$\alpha_T^{(W)} / (\pi T) \to 1$} and the conjectural bound {\eqref{eq:conjecured_bound}} becomes equivalent to the universal one {\eqref{eq:boundonchaos}}. This equivalence suggests further the plausibility of the conjecture~\eqref{eq:conjecured_bound}. 

\subsection{SYK Model}\label{sec:SYK_finiteT}
  To illustrate the foregoing discussion, and provide some evidence for the hypothesis at finite-$T$~\eqref{eq:hypothesis_finite_temp} and the conjectural bound on chaos \eqref{eq:conjecured_bound}, let us consider again the example of SYK model. 
   
  At low temperatures $T = 1/\beta \ll J$, it is well-known that {$\lambda_{L,T} = 2 \pi T$}~\cite{kitaev15} saturates the universal quantum bound \eqref{eq:boundonchaos}. In this limit, the finite-$T$ autocorrelation function of $\O = \sqrt{2} \gamma_1$ may be computed exactly by conformal invariance~\cite{sykcomment}. Choosing the Wightman inner product, we have
 \begin{equation} C_{\beta}^{W}(t) \propto \sech \left( t \pi T \right)^{2/q} \,. \end{equation}
This is the autocorrelation function of the exact solution \eqref{eq:universal_wavefunction_main_text}, and corresponds to Lanczos coefficients $b_{n,T}^{(W)} = \pi T \sqrt{n(n-1+\eta)}$. They satisfy the hypothesis~\eqref{eq:hypothesis_finite_temp} with $\alpha_T^{(W)} = \pi T$ \eqref{eq:alpha_lowT}. Therefore the low-temperature SYK model saturates also our {conjectural bound \eqref{eq:conjecured_bound}}. 
	
\begin{figure}
	\centering
	\includegraphics{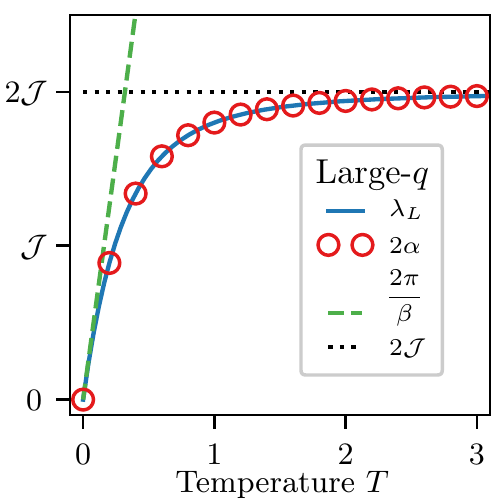}
	\caption{Exact Lyapunov exponent {$\lambda_L(T)$} \eqref{eq:lambda_SYK_q} and growth rate {$\alpha(T)$} with the Wightman inner product \eqref{eq:alpha_SYK_q} of the SYK model in the large-$q$ limit as a function of temperature (in units of coupling constant $\mathcal{J}$). The conjectured bound {$\lambda_L(T) \le 2 \alpha(T)_W$} is exactly saturated at all temperatures, while the universal bound {$\lambda_L(T) \le 2 \pi T$} only saturates in the zero temperature limit.}
	\label{fig:SYK_finite_temp}
\end{figure}

At finite (but not necessarily low) temperatures, using analytic results in the large-$q$ limit~\cite{sykcomment}, it is not hard to check (see Appendix~\ref{app:SYK}) that our conjectured bound \eqref{eq:conjecured_bound} is saturated, whereas the universal bound~\eqref{eq:boundonchaos} is not, see Fig.~\ref{fig:SYK_finite_temp}. This result indicates that an extension of our bound on chaos to finite temperature is at least plausible. The exact agreement between {$\alpha_{T}^{(W)}$ and $\lambda_{L,T}$} is notable given that the former is defined solely from 2-point correlators whereas the latter requires 4-point functions. 
	
We reiterate that the above SYK results depend crucially on the Wightman inner product. If the ``standard'' inner product \eqref{eq:inner_finite_T_CMT} is chosen instead, the Lanczos coefficients {$b_{n,T}^{(S)}$} cannot be extracted from the conformal solution, since that would require the Taylor expansion of $C_{\beta}^S(t)$ around $t=0$, at which the conformal solution is {non-analytic}. A  numerical high-temperature expansion (extending the method of Appendix~\ref{app:SYK}) and an exact calculation in the large-$q$ limit both indicate that the Lanczos coefficients still grow linearly, but the growth rate \textit{increases} as the temperature decreases.

To summarize, exact calculations in the SYK model support the universal operator growth hypothesis at finite temperature, and the conjectural bound on chaos. 

%To conclude this section, we examine the changes resulting from another choice of inner product. 

	\section{Conclusions}\label{sec:discussion}
	
	\subsection{Discussion}
	
	We have presented a hypothesis on the universal growth of operators: the Lanczos coefficients follow the asymptotically linear form $b_n = \alpha n + \gamma + \smallo(1)$ in non-integrable systems, with a logarithmic correction in 1d. We have seen copious evidence that the hypothesis is satisfied in a wide variety of non-integrable models.  Over the course of this work, the growth rate $\alpha$ has emerged as a quantity of prime importance, tying a diverse array of seemingly-disparate ideas together. %Indeed, this work functions almost a field guide to the difference guises $\alpha$ takes and how to translate between them. 
	Let us recount them now: 
	\begin{itemize}
		\item $\alpha > 0$ is the slope of asymptotically linear growth of the Lanczos coefficients.
		\item $\tfrac{2}{\pi}\alpha = \omega_0$ is the exponential decay rate of the spectral function $\Phi(\omega) \sim e^{-|\omega|/\omega_0}$, {which can be (and has been) measured experimentally~\cite{PhysRev.188.609,PhysRevB.10.822,0953-8984-2-50-017}}.
		\item $\pm i \pi / (2\alpha)$ are the locations of the singularities closest to the origin in the (analytic continuation) of the autocorrelation $C(t)$, see Appendix~\ref{app:recursion-review}.
		\item $2\alpha$ is the exponential growth rate of  Krylov-complexity.
		\item $2 \alpha$ is an upper bound for the growth of all q-complexities.
		\item $2\alpha$ is an upper bound for the Lyapunov exponent (whenever the latter is well-defined), since quantum OTOCs are an example of q-complexities.
	\end{itemize}
	We have, of course, put aside the precise conditions and qualifiers of each statement. In light of these results, $\alpha$ plays a central role in operator growth and dynamics of complex systems.

	Complexity --- especially the Krylov-complexity --- arose as a key concept in this work. We would like to highlight its temporal nature which, as we now argue, makes it a more general notion than chaos. Chaos essentially tracks the development of structures at ever-smaller scales in \textit{phase space}. In classical systems, of course, this may proceed indefinitely, while in quantum systems, features smaller than $\hbar$ are ruled out and the process saturates. Chaos therefore cannot carry over straightforwardly to systems deep in the quantum regime, where the phase space volume is comparable to $\hbar$ and saturation occurs almost immediately. The K-complexity, in sharp contrast, measures structures at ever-smaller scales in the \textit{time} domain.  We believe this is a fundamental difference; as we have seen, the K-complexity can grow exponentially in quantum systems beyond semiclassical or large-$N$ limits. Operator complexity may well supersede the notion of chaos in quantum dynamics.

	\subsection{Outlook}
	
	We would like to understand how our hypothesis can be affected by obstructions to thermalization. Based on evidence available to us, it is tempting to conjecture that they lead to a qualitative slower growth for quantum systems. Confirming this in general would be a remarkable result. However, given the diversity of non-thermalizing situations, it may be more reasonable to explore them on a case by case basis. 
	In free and integrable models, there are an extensive number of conserved local or quasi-local charges. The behavior of the Lanczos coefficients in integrable models is likely non-universal, and depends strongly on the model and operator in question~\cite{viswanath2008recursion}. We wish to gain general analytical insights in this direction (especially for interacting models), by leveraging the knowledge available on the quantum inverse scattering method~\cite{Faddeev1999,kitanine1999647,maillet2000627}. Also, it may be desirable to modify the Lanczos algorithm to promote the semi-infinite line to a lattice where the perpendicular direction is generated by commutators against quasi-local conserved charges. 
	Another exceptional case is quantum scar states~\cite{scar1,scar2,scar3}, isolated states that fail to thermalize in otherwise chaotic systems, possibly due to emergent or approximately conserved charges. It would be revealing to see how scars are reflected in the Lanczos coefficients.
	Finally it would be of great interest to understand the interplay of the hypothesis with many-body localized systems (see \cite{abanin2018mbl} and references therein for a review, and \cite{khait16mbl} for numerical calculations of Lanczos coefficients in disordered spin chains) where thermalization fails.
	
  Our treatment at finite temperatures is far from complete and leaves numerous open questions, especially those concerning the ``standard'' inner product: How do the Lanczos coefficients grow? If linearly, how does the growth rate depend on the temperature? How can we extend our bound on chaos to finite $T$? Numerical investigations into these questions are challenging due to the presence of the thermal density matrix~\cite{lucas2018operator, auerbach2018hall,martyn2018product}. Quantum Monte Carlo seems promising for this problem, as the Lanczos coefficients can be computed without analytic continuation. In low dimensions, DMRG can be also useful: matrix product operators can be used to approximate the thermal state, and the operators in the Lanczos algorithm.% Extending our bound on chaos to finite temperature is another key 
	
	One would like to put the hypothesis on more solid mathematical footing, especially in 1d. Finding analytically tractable models \textit{far from the large-$N$ limit} that achieve the maximal Lanczos coefficient growth seems a formidable problem, which is made even harder by the restriction to {time-independent Hamiltonian} systems; the only result in this direction is that of \cite{bouch2015complex} in 2d. Many solvable models of quantum chaos (see Refs~\cite{chan2018solution,prosen1} for notable recent progress) are only defined as unitary maps or Floquet systems. To this respect, a meaningful extension of the hypothesis to such contexts would be a highly rewarding advance.
	
	An alternative route would be to develop an extended (Hermitian) random matrix theory. Standard proofs of the Wigner semicircle law exploit the connections between the moments of a distribution, the combinatorics of Dyck paths, Catalan numbers, and the Stieltjes transform of a distribution~\cite{tao2012topics}. These are directly analogous to the moments $\mu_{2n}$, the combinatorics of Motzkin paths, secant numbers, and the continued fraction expansion for $G(z)$ --- all of which arose in the calculation of our exact wavefunction in Appendix~\ref{app:universal}). The non-trivial appearance of the same type of objects in both contexts suggests a strong analogy. We thus speculate that the hypothesis can be derived analytically by introducing a new type of random matrix ensemble that incorporates locality and translation invariance. (This is similar to the framework of \cite{PhysRevLett.107.097205}.) In this case, a Hamiltonian such as $H = \sum_{<x,y>} h_{x,y}$, where $h_{x,y}$ is a random matrix acting on neighboring sites $x$ and $y$, should obey the hypothesis \eqref{eq:linear_lanczos} in expectation. Therefore generic, 2-local Hamiltonians would also be expected to obey the hypothesis by concentration of measure. It may well be that showing the hypothesis holds for a specific Hamiltonian is of comparable difficulty to showing the ergodic hypothesis applies to specific classical systems. 
	
  Coming back to physics, we argue that there should be a general principle, analogous of the second law of thermodynamics, that governs the operator growth in generic systems. Indeed, the latter is \textit{irreversible}, in the same sense as the dynamics of an isolated gas is so in the thermodynamic limit. We cannot help but wonder what \textit{entropy} is maximized by the operator growth process, and whether any notion of (quantum) dynamical entropy (see e.g. \cite{connes1987,alicki-fannes,benatti2012deterministic,prosen2007chaos} is relevant in describing the process. Elusive as it seems, such a thermodynamic principle might be the ultimate explanation of our empirical observations of ubiquitous maximal operator growth.
	
   To close, we wish to point out that the territory of q-complexities beyond K-complexity and OTOCs is completely unexplored. In generic many-body systems (i.e. not semiclassical) at infinite temperature, these two examples represent two extremes, showing maximal and non-existent exponential growth rates, respectively.  The significant gap between them should be filled with potentially more meaningful measures of complexity. These complexities could be 
	entirely new concepts or disguised forms of existing notions such as circuit complexity and entanglement entropy. Hopefully, charting this \textit{terra incognita} will continue to shed new light on the complex nature of many-body quantum dynamics.

	\textit{Acknowledgments}. 
	We thank Assa Auerbach, Erez Berg, David Huse, Joel Moore, Steve Shenker, and Achim Rosch for insightful discussions. Part of the numerical computations are performed using the cluster of Laboratoire de Physique Th\'eorique et Mod\`eles Statistiques (CNRS, Universit\'e Paris-Sud). We acknowledge support from NSF Graduate Research Fellowship Program NSF DGE 1752814 (DP), the Emergent Phenomena in Quantum Systems initiative of the
	Gordon and Betty Moore Foundation (TS), ERC synergy Grant UQUAM (EA, XC and TS) and DOE grant DE-SC0019380 (EA and XC).
	
	\bibliography{references}

	\appendix
	
	\section{Brief Review of the Recursion Method}
	\label{app:recursion-review}
	In this appendix we recall the relations between Lanczos coefficients,  correlation function, Green function, spectral function, and moments. These relations are mathematical in nature, and apply to any inner product on the operator space, and thereby to finite as well as infinite temperature. For simplicity, we will omit the sub- and superscripts indicating the inner product.

	Let us recall the five equivalent representations of the dynamics of an operator:
	\begin{equation}
		C(t) \leftrightarrow G(z) \leftrightarrow \Phi(\omega) \leftrightarrow \{\mu_{2n}\} \leftrightarrow \{b_n\}.
	\end{equation}
	The first four are related by linear transformations given in the text. For instance, the \textit{moments} $\mu_{2n}$ are  the Taylor expansion coefficients of autocorrelation around $t = 0$:
	\begin{equation}
	C(-it) := \sum_{n=0}^\infty \mu_{2n} \frac{t^{2n}}{(2n)!}, \quad \mu_{2n} := \pbraket{\O|\L^{2n}|\O},
	\label{eq:moments}
	\end{equation}
	where the odd terms vanish provided $\O$ is Hermitian. The moments can also be extracted from the spectral function via
	\begin{equation}
	    \mu_{2n} = \int \omega^{2n} \Phi(\omega) \, d\omega.
	\end{equation}
	All the transformations between the first four quantities are similarly straightforward. 
	
	The Lanczos coefficients, on the other hand, are related to the others via a \textit{non-linear} transformation. The rest of this Appendix discusses how to perform the non-trivial translation between the Lanczos coefficients and the moments both asymptotically and numerically.

	\subsection{From Moments to Lanczos Coefficients}
	Cumulative products of the first $n$ Lanczos coefficients are given by determinants of the Hankel matrix of moments~\cite{viswanath2008recursion}
	\begin{equation}
	b_1^2 \dots b_n^2 =  \det \left( \mu_{i + j} \right)_{0 \leq i,j \leq n}  \,.
	\end{equation}
	If the moments are known, the determinant can be computed efficiently by transforming the Hankel matrix into diagonal form. Doing this iteratively for $k \in [1,n]$ provides a fast algorithm that computes $b_1,\dots, b_n$ from $\mu_{2},\mu_{4},\dots,\mu_{2n}$. The algorithm may be expressed concisely as a recursion relation (see Eq. 3.33 of Ref.~\cite{viswanath2008recursion}) as follows:
	\begin{align}
	&  b_n = \sqrt{M_{2n}^{(n)}} \,, \nonumber \\
	&  M_{2k}^{(0)} = \frac{M_{2k}^{(m-1)}}{b_{m-1}^2} -
	\frac{M_{2k-2}^{(m-2)}}{b_{m-2}^2} \,,\, k=m,\dots,n \,, \nonumber \\ 
	&     M_{2k}^{(0)} = \mu_{2k} \,,\, b_{-1} = b_{0} := 1\,,\, M_{2k}^{(-1)} := 0 \,.
	\label{eq:recipe}
	\end{align}
	If an analytic expression for $C(t)$ is known, then an arbitrary number of the Lanczos coefficients may be computed numerically via \eqref{eq:recipe}. We remark that this algorithm suffers from large numerical instabilities due to repeated floating-point divisions.
	
	\subsection{From Lanczos Coefficients to Moments}\label{app:recursion_2}
	It follows from the tridiagonal form of $L$ that the moments may be expressed in terms of the Lanczos coefficients as
	\begin{equation}
	\mu_{2n} = \pbraket{\O|\L^{2n}|\O} = (L^{2n})_{00}.
	\end{equation}
	If the Lanczos coefficients are known, this is a completely combinatorial object. In particular, the moments are given by a sum over \text{Dyck paths}. Formally, a Dyck path of length $2n$ can be defined as a sequence $(h_0, h_1, \dots, h_2n)$ such that:  $h_0 = h_{2n} = 1/2$; $h_k \geq \frac12$ and $|h_k - h_{k+1}| = 1$ for any $k$. These are often visualized as paths starting at height zero where each segment either increases or decreases the height by one unit, with the constraint that the height is always non-negative and returns to zero at the end. Denoting the set of such paths by $\mathcal{D}_n$, we have
	\begin{equation} \label{eq:Dyck}
	\mu_{2n} = \sum_{\st{h_k} \in \mathcal{D}_n} \prod_{k=1}^{2n} b_{(h_k + h_{k-1})/2} \,.
	\end{equation}
	For example, $\mu_2 = b_1^2$ and $\mu_4 = b_1^4 + b_1^2 b_2^2$. The number of Dyck paths of length $2n$ is given by the Catalan numbers $C_n = \frac{(2n)!}{(n+1)!n!}$.
	A consequence of \eqref{eq:Dyck} is the following lower bound:
	\begin{equation}
	\mu_{2n} \geq b_1^2 \dots b_n^{2} \,. \label{eq:moment_bn_bound}
	\end{equation}
    On the other hand, we have the upper bound $\mu_{2n} \le \max_{k=1}^n \left(b_k^2\right) C_n$.
%	Since $C_n \sim \BigO(4^n n^{-3/2})$ grows exponentially, the two bounds imply that the moments and the cumulative products of Lanczos coefficients $\prod_{k=1}^n b_k^2$ have growth rates that differ at most by an exponential in $n$. 
	 Applying the upper and lower bounds, linear growth of the Lanczos coefficients $b_n$ corresponds to the following growth rate of moments:
	\begin{equation}
	\mu_{2n} = \exp( 2 n \ln n +\BigO(n)) \,.   \label{eq:moments_growth}
	\end{equation}
	This equation is a useful reformulation of the linear growth hypothesis.
	
    If the growth rate is known as well, $b_n = \alpha n + \BigO(1)$, it is possible to refine the asymptotic by specifying the next order exponential term:
	\begin{equation}
	\mu_{2n} = \left( \frac{4 n \alpha}{e \pi} \right)^{2n} e^{\smallo(n)} \,.  \label{eq:moment_growth1}
	\end{equation}
	Combining this equation with the Stirling formula, the correlation function $C(t) = \sum_n \mu_{2n} (it)^{2n} / (2n)!$ has convergence radius $r = \pi / (2 \alpha)$, due to singularities at $t = \pm i r$; in fact, $C(t)$ is analytical in the strip $-r < \operatorname{Im} (t) < r $, see Fig.~\ref{fig:strip}.  Therefore, the Fourier transform of $C(t)$, which is the spectral density $\Phi(\omega)$, has a exponential decay
	\begin{equation}\label{eq:alpha_omega0_app}
	\n{\Phi(\omega)} =  e^{-|\omega|/\omega_0 + o(\omega)} \,,\, \omega_0 = r^{-1} =  2 \alpha / \pi \,.
	\end{equation}

	We illustrate the above results by a simple example: when $b_n  = \alpha\,  n$, then $C(t) = \sech(\alpha t)$ and $\Phi(\omega) =  \frac{\alpha}{\pi} \sech\left( \frac{\pi \omega} {2 \alpha}\right)$. The moments $\mu_{2n} = 1,1,5,61,1385,\dots$ are known as Euler or secant numbers and have the asymptotic behavior $\mu_{2n} = 4 \sqrt{\frac{4n}{\pi}} \left( \frac{4n}{\pi e}\right)^{2n} (1 + \smallo(1))$~\cite{borwein1989pi}. We checked that \eqref{eq:alpha_omega0_app} and ~\eqref{eq:moment_growth1} hold in all analytic examples we are aware of in the literature and believe them to hold in general.  
	
	\section{Moments and Lanczos Coefficients in the SYK Model}
	\label{app:SYK}
	In this section we compute the Lanczos coefficients in the large-$N$ SYK model at infinite temperature with the initial operator $\O = \sqrt{2} \gamma_1$. Most often, this is done by computing the moments and applying the mapping described in Section~\ref{app:recursion-review}.
	
	For convenience, we recall the SYK Hamiltonian and disorder normalization: 
	\begin{align}
	H_\text{SYK}^{(q)} &= i^{q/2} \sum_{1\leq i_1 < i_2 < \dots < i_q \leq N} J_{i_1\dots i_q}
	\gamma_{i_1}  \cdots \gamma_{i_q},\, \\ 
	\overline{J_{i_1\dots i_q}^2} &= 0,\\
	\overline{J_{i_1\dots i_q}^2}^2 &= \frac{(q-1)!}{N^{q-1}} J^2,
	\end{align}
	where the line denotes disorder averages. We shall extend $J_{i_1\dots i_q}$ to all $i_1, \dots, i_q$ by anti-symmetry. As discussed in the main text, disorder-averaging will be assumed throughout. We first describe the general method, and then discuss the large-$q$ limit. 
	
	\subsection{General Method}
	Since the moments are closely related to the Green function, they can be calculated by the diagrammatic technique commonly used in the SYK literature. Indeed, $\mu_{2n}$ can be represented as a sum over diagrams $G$ diagrams with $2n$ vertices:
	\begin{equation}
	\mu_{2n} = J^{2n} 2^{(2-q)n} \sum_{G} C_G  \,,
	\end{equation}
	where $C_G$ is the combinatorial factor of the diagram, which counts the number of labellings of the vertices by $1,\dots,2n$ such that the labels are increasing from left to right.
	
	Let us illustrate the diagrams with some examples with $q=4$ and $n=1,2$. Direct calculation yields:
	\begin{equation}
	\begin{aligned}
	\L \gamma_1 &= -\sum_{j<k<l} J_{1jkl}\gamma_{j} \gamma_k \gamma_{l} \,, \\
	\L^2 \gamma_1 &= 2^{2-q} \sum_{j<k<l} J_{1jkl}^2 \gamma_1 \\
	&\quad+ \sum_{j<k<l} J_{1jkl} \sum_{r<s<t} J_{jrst} 
	\gamma_{k} \gamma_l \gamma_r \gamma_s \gamma_t   \\ 
	&\quad+ \sum_{j<k<l} J_{1jkl} \sum_{r<s<t} J_{krst} 
	\gamma_{j} \gamma_l \gamma_r \gamma_s \gamma_t  \\
	&\quad + \sum_{j<k<l} J_{1jkl} \sum_{r<s<t} J_{lrst} 
	\gamma_{j} \gamma_k \gamma_r \gamma_s \gamma_t \,.  
	\end{aligned}
	\end{equation}
	The first two moments $\mu_2$ and  $\mu_4$ are (twice) the norm squared of the $\L \gamma_1$ and $\L^2 \gamma_1$, respectively. Under disorder averaging, the terms on the right-hand side are orthogonal, and each corresponds to a different diagram:
	\begin{equation}
	\begin{aligned}
	\mu_2 =  J^{2} 2^{(2-q)} \ &=\  \includegraphics[valign=c]{drawing-1.mps} \,, \\
	\mu_4 = J^4 2^{2(2-q)} q  \ &=\      \includegraphics[valign=c]{drawing-2.mps}\\
	&\quad +\ \includegraphics[valign=c]{drawing-3.mps}\\
	&\quad +\ \includegraphics[valign=c]{drawing-4.mps}\\
	&\quad +\ \includegraphics[valign=c]{drawing-5.mps}.\\
	\end{aligned}
	\end{equation}
	The combinatorial factor is $C_G = 1$ for each of the above graphs. The first non-trivial combinatorial factor is$C_G =6$ for the diagram \includegraphics[valign=c]{drawing-6.mps}, which contributes to $\mu_6$. The six vertex orderings are $1 \; \begin{matrix} 2 & 3 \\ 4 & 5 \end{matrix} \; 6$, $1 \; \begin{matrix} 4 & 5 \\ 2 & 3 \end{matrix} \; 6$, $1 \; \begin{matrix} 2 & 4 \\ 3 & 5 \end{matrix} \; 6$, 
	$1 \; \begin{matrix} 3 & 4 \\ 2 & 5 \end{matrix} \; 6$, $1 \; \begin{matrix} 2 & 5 \\ 3 & 4 \end{matrix} \; 6$, and $1 \; \begin{matrix} 3 & 5 \\ 2 & 4 \end{matrix} \; 6$. 
	
	The SYK diagrams encode the Schwinger-Dyson equations governing the autocorrelation and Green's function which are, up to trivial transformations, the exponential and ordinary generating functions of the moments, respectively:
	\begin{subequations}
		\begin{align}
		z G(z) &= 1 + J^2 2^{2-q} G(z) \widetilde\Sigma(z),\\
		\Sigma(t) &= C(t)^{q-1}, \label{eq:SD1}\\\ 
		\widetilde\Sigma(z) &= i\int_0^{\infty} \Sigma(t) e^{-itz} dt,
		\end{align}
		\label{eq:SD}
	\end{subequations}
	that is, $\widetilde\Sigma(z)$ and $\Sigma(t)$ are related by (non-standard) Laplace transform \eqref{eq:laplace_transform} just as $G(z)$ and $C(t)$ are. Equation \eqref{eq:SD} can be represented diagrammatically (here for the case $q=4$) by
	\begin{equation}
	\includegraphics[valign=c]{drawing-9.mps} = 
	\includegraphics[valign=c]{drawing-8.mps} + \includegraphics[valign=c]{drawing-7.mps} \,.
	\end{equation}
	The dot represents a general SYK diagram (a fully-dressed Green's function). This is the sum of the bare Green's function, or the time-domain product of $(q-1)$ dressed Green's functions.  Note that both exponential and ordinary generating functions are needed to take the combinatorial factors into account: a serial (respectively, parallel) composition of diagrams correspond to product of ordinary (resp. exponential) generating function. 
	
	Equation \eqref{eq:SD} has no closed form solution for general $q$. However, working with the power series representations, it enables the numerical calculation of $\mu_{2}, \dots, \mu_{2n}$ in polynomial time and space complexity in $n$. Concretely, the following iteration algorithm can be easily implemented in a computer algebra system:
	\begin{enumerate}
		\item Set $g_0(z) := z^{-1}$, and let $j = 0$.
		\item Compute $c_j(t)$ from $g_j(z)$ by replacing $z^{-2n-1}$ with $(it)^{2n}/(2n)!$. 
		\item Set $\sigma_j(t) := c_j(t)^{q-1}$ up to order $t^{j}$. 
		\item Compute $\widetilde \sigma_j(z)$ from $\sigma_j(t)$ by replacing $(it)^{2n}$ with $z^{-2n-1} (2n)!$.
		\item Set $g_{j+1}(z) := (1 + J^2 2^{2-q} g(z) \widetilde\sigma_j(z))/z$ up to order $t^{j+1}$.
		\item Increment $j$ by $1$ and repeat from step 2.
	\end{enumerate}
	When the above procedure is stopped at $j = n$, the result $g_n(z)$ will be a polynomial truncation of the Green function: $g_{n}(z) = \sum_{j=0}^{n} \mu_{2j} z^{-2j-1}$, which contains the correct moments up to $\mu_{2n}$. They can be then used to compute Lanczos coefficients $b_1^2, \dots, b_n^2$  by the recipe \eqref{eq:recipe}. Arbitrary-precision rational number arithmetic is necessary for $n\sim 10^2$, since the moments grow very fast. We calculated $b_n$ for a few different values of $q$ up to $n=100$, and extracted the linear slope by a linear fit. The results are reported in Table~\ref{tab:syk_exponents} and Fig.~\ref{fig:linear_lanczos_evidence} (a).
	
\begin{figure}
	    \centering
	    \includegraphics{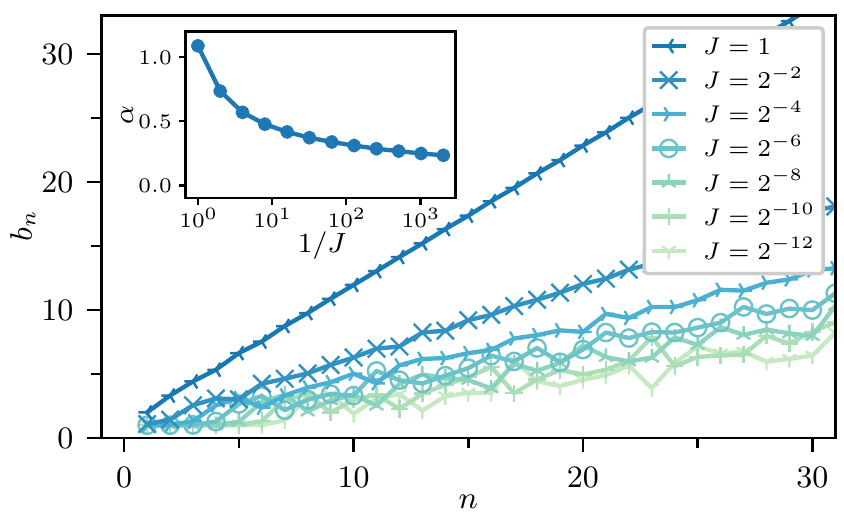}
	    \caption{Change in the growth rate near integrability for the SYK model with $q=2$ and $q=4$ \eqref{eq:SYK24}. The ratio of the $q=4$ to $q=2$ term is given by $J$, and the model becomes free at $J=0$.}
	    \label{fig:chaos_transition_SYK}
	\end{figure}

    The above method can be readily adopted to variants of SYK where two-body and four-body interactions coexist:
	\begin{equation}
	    H = H_\text{SYK}^{(4)}(J) + H_{\text{SYK}}^{(2)}(J = 1) \,. \label{eq:SYK24}
	\end{equation}
	One only needs to replace the last term in \eqref{eq:SD1} by a sum over $q=2$ and $q=4$ with the corresponding coupling constants. Since the $q=2$ model is non-interacting, eq.~\eqref{eq:SYK24} can be another model to study the effect of weak thermalizing interaction on the Lanczos coefficients. The results, shown in Fig.~\ref{fig:chaos_transition_SYK}, are qualitatively consistent with those from the Ising model (Fig.~\ref{fig:linear_lanczos_evidence}): the linear growth rate depends only weakly on the interaction strength $J$ as it goes to zero. Quantitative, a logarithmic dependence 
	\begin{equation} \alpha \sim 1/\ln (1/J) \end{equation} 
	describes the numerical data well for vanishing $J$.
	
	\subsection{Large-$q$ limit}
	In the large-$q$ limit, \eqref{eq:SD} can be solved analytically. It is convenient to define the coupling constant~\cite{sykcomment,roberts2018operator}
	\begin{equation}
	\mathcal{J}^2 := 2^{1-q}\, q  \, J^2 \,.
	\end{equation}
	It is then known \cite{sykcomment,roberts2018operator} that $C(t)$ admits a $1/q$ expansion
	\begin{align}
	C(t) = 1 + \frac1q \mathcal{C}(t) + \BigO(1/q^2) \,,\,
	\end{align}
	where the leading non-trivial term satisfies the following differential equation:
	\begin{align}
	\mathcal{C}''(t) = - 2 \mathcal{J}^2 e^{\mathcal{C}(t)} \,,\, \mathcal{C}(0)= \mathcal{C}'(0) = 0 \,,
	\end{align}
	whose solution is
	\begin{equation}\label{eq:qsyk_Ct}
	C(t) = 1 +\frac2q  \ln \sech(\mathcal{J } t) + \BigO(1/q^2) \,.
	\end{equation}
	The corresponding moments
	\begin{align}
	\mu_{2n} = \frac{2}{q} \mathcal{J}^{2n} T_{n-1} + \BigO(1/q^2)   \,,\, n > 0 \,,
	\end{align}
	where $(T_n)_{n=0}^{\infty} = (1, 2, 16, 272, 7936,\dots)$ are the tangent numbers~\cite{oeis_tangent}. The generating function of $T_n$ admits a  continued fraction expansion~\cite{oeis_tangent}:
	\begin{equation} \sum_{n=0}^{\infty} T_n x^{n} = \dfrac{1}{1- \dfrac{1\times 2 x}{1-\dfrac{2 \times 3x}{1- \dfrac{3\times 4x}{1-\ddots}}}}
	\end{equation}
	Using this, one can obtain the following Lanczos coefficients for the large-$q$ SYK model
	\begin{equation}
	b_n^{\text{SYK}} = \begin{dcases} 
	\mathcal{J}\sqrt{2/q} + \BigO(1/q) & n = 1  \\
	\mathcal{J} \sqrt{n(n-1)} + \BigO(1/q) &  n > 1 \,.
	\end{dcases}
	\label{eq:SYK_Lanczos_coeffs}
	\end{equation}
	It is not hard to check using \eqref{eq:1D_chain_problem} that the wavefunction on the semi-infinite chain is
	\begin{equation}
	\varphi_n(t) =
	\begin{dcases}
	1 +\frac2q  \ln \sech(\mathcal{J } t) + \BigO(1/q^2)  & n= 0 \\
	\tanh(\mathcal{J } t)\sqrt{\frac{2}{nq}} + \BigO(1/q^2)  & n > 0 \,.
	\end{dcases} \label{eq:sykqvarphi}
	\end{equation}
	The corresponding probability distribution is identical to the operator size distribution (see Eq. (5.11) of Ref.~\cite{roberts2018operator}):
	\begin{equation}
	P_{s} (t) =  |\varphi_n(t)|^2 \,,\, s = 1 + n (q-2) \,. \label{eq:compare_syk}
	\end{equation}
	
	The large-$q$ SYK model is also solvable at any finite temperature~\cite{sykcomment}. The temperature $T$ is parametrized by $v \in (0,1)$ via
	\begin{equation} \frac{T}{\mathcal{J}} =\frac{ \cos\frac{\pi v}2 }{\pi v}  \label{eq:def_v_sykq}  \,. \end{equation}
	The limits $T \to \infty$ and $T\to 0$ correspond to $v \to 0$ and $v\to 1$, respectively. The Lyapunov exponent is then
	\begin{equation} \lambda_{L,T} = 2 v \pi T  \,, \label{eq:lambda_SYK_q} \end{equation}
	and the autocorrelation under the Wightman inner product~\eqref{eq:inner_finite_T} is 
	\begin{equation} C_\beta^{W}(t) = 1 + \frac2q \ln \sech\left( v t \pi T \right) + \BigO(1/q^2) \,. \end{equation}
	Comparing to \eqref{eq:qsyk_Ct}, we see immediately that
	\begin{equation} 
	b_{n,T}^{(W)} = \begin{dcases} 
	v  \pi T \sqrt{2/q} + \BigO(1/q) & n = 1  \\
	v   \pi T \sqrt{n(n-1)} + \BigO(1/q) &  n > 1 \,.
	\end{dcases}\label{eq:alpha_SYK_q}
	\end{equation}
	Therefore the finite-$T$ growth rate with the Wightman inner product is \begin{equation} \alpha_T^{(W)} = v \pi T  \label{eq:alphaW_sykq} \end{equation}
	at any temperature. Thus, the bound $\lambda_{L,T}\le 2 \alpha_T^{(W)}$ is saturated at all temperature in the SYK model, whereas the bound $\lambda_{L,T} \le 2 \pi T$ is only so in the zero-temperature limit (see Fig.~\ref{fig:SYK_finite_temp}). Using the relation between growth rate and spectral function decay rate \eqref{eq:spectral_density_asymptotic} and the relation~\eqref{eq:standard_vs_Wightman} between spectral functions of different inner products, it is not hard to obtain the growth rate with the standard inner product from \eqref{eq:alphaW_sykq}:
	\begin{equation}
	    \alpha_{T}^{(S)} = \frac{v \pi T}{1-v} \,.
	\end{equation}
	Using \eqref{eq:def_v_sykq}, we obtain the limits $\alpha(T)_S \to \mathcal{J} \pi / 2$ as $T \to 0$ and $\alpha_T^{(S)} \to  \mathcal{J} $ as $T \to \infty$. We notice that $\alpha(T)_S$ \textit{increases} at low temperatures while, in contrast, $\alpha_T^{(W)}$ decreases.

	\section{Numerical Details for 1d Spin Chains}
	\label{app:numerics}

	This section discusses the numerical details involved in computing the Lanczos coefficients and Krylov basis vectors in 1D spin chains. We work directly in the thermodynamic limit of a chain with $N \to \infty$ sites. However, bookkeeping will reduce this to finite-dimensional matrix multiplication. 
	
	Suppose we have a translation-invariant $k$-local Hamiltonian $H = \sum_n h_n$ and an $\ell$-local operator $\O = \sum_n \O_n$. Here $h_n$ and $\O_m$ are operators starting on sites $n$ or $m$ respectively. (For instance, we might have $\O_2 = \cdots \otimes I_1 \otimes X_2 \otimes Z_3 \otimes I_4 \otimes \cdots$.) We normalize the operators so that $\pbraket{h_n|h_n} = 1 = \pbraket{\O_m|\O_m}$. At minor additional computational cost, we can work with an operator at a finite wavevector $q$:
	\begin{equation}
	\O_q = \sum_n \O_n e^{iqn}.
	\label{eq:wavevector_q_operator}
	\end{equation}
	The crucial point is that applying the Liouvillian to $\O_q$ is another operator at wavevector $q$ by using translation-invariance to re-index the sum at the cost of phase factors. Explicitly,
	\begin{equation}
	[H,\O_q] = \sum_{m,n} [h_n, \O_m] e^{iqm} = \sum_m \O_m' e^{iqm}
	\label{eq:commutator_q_wavevector}
	\end{equation}
	where
	\begin{equation}
	\O_m' = \sum_{n=m-k+1}^{m-\ell+1} e^{iq s_{nm}} [h_{n+s_{nm}},\O_{m+s_{nm}}] 
	\label{eq:new_q_wavevector_operator}
	\end{equation}
	where the shift is $s_{nm}$ is the index of the first non-identity site of $[h_n,\O_m]$ minus $m$, which is needed to keep track of how much the support of the operator shifted due to the commutator. One can check that $\O_m'$ starts on site $m$.
	
	Therefore we only need to keep track of operators starting on a single site, say site $0$. We adopt the basis of Pauli strings and, following, e.g. \cite{dehaene2003clifford}, we adopt a representation which minimizes the computational cost of taking commutators. Since $iY = ZX$, we may adopt a representation
	\begin{equation}
	i^\delta (-1)^\epsilon Z_1^{v_1} X_1^{w_1}\otimes \cdots \otimes Z_n^{v_n} X_n^{w_n}
	\end{equation}
	where $\delta, \epsilon, v_k, w_k \in \st{0,1}$, i.e. a Pauli string of length $n$ may be represented by two binary vectors $\v{v}$ and $\v{w}$ of length $n$ and two binary digits. So if $\tau_1 = i^{\delta_1} (-1)^{\epsilon_1} Z^{\v{v}_1} X^{\v{w}_2}$ and $\tau_2 = i^{\delta_2} (-1)^{\epsilon_2} Z^{\v{v}_2} X^{\v{w}_2}$, then their commutator is a string $\tau' = [\tau_1, \tau_2]$ with 
	\begin{equation}
	\begin{aligned}
	\delta' &= \delta_1 + \delta_2,\\
	\epsilon' &= \epsilon_1 + \epsilon_2 + \delta_1 \delta_2 + \v{w}_1 \cdot \v{v}_2,\\
	\v{v}' &= \v{v}_1 + \v{v}_2,\\
	\v{w}' &= \v{w}_1 + \v{w}_2.
	\end{aligned}
	\label{eq:pauli_mult_efficient}
	\end{equation}
	All additions are performed over $\Z_2$. 
	
	With this setup, the Lanczos coefficients can be computed in a similar way to matrix-free exact diagonalization codes. A translation-invariant operator can be stored as a hash map of Pauli strings starting on site zero with complex coefficients. The Liouvillian is applied by combining \eqref{eq:commutator_q_wavevector}, \eqref{eq:new_q_wavevector_operator}, and \eqref{eq:pauli_mult_efficient}. Of course, it is not necessary to take $\O$ to be translation invariant. One could equally well take a small single-site operator and apply the same technique without the sum over all sites. We note that the Lanczos algorithm \eqref{eq:Lanczos_algorithm} only requires the storage of three operators at any time. In practice the method described here allows a few dozen Lanczos coefficients to be computed in a few minutes on a modern laptop and is generally memory-limited by the exponential increase in the number of Pauli strings required.
	
	Once the Lanczos coefficients and Krylov vectors have been computed, it is possible to understand how the operators $\O_n$ grow in physical space. One way to characterize this is in terms of the distribution of string lengths in each $\O_n$. If $\O_n = \sum_{a} c_a \sigma^{a}$, where the sum runs over all Pauli strings $a$, then the distribution is defined by $P_n(s) = \sum_{a : \n{a} = s} \n{c_a}^2$. This distribution is shown for the Hamiltonian $H_1$ with the parameters given in Fig. \ref{fig:linear_lanczos_evidence}. The mean and variance of the distribution grow with $n$. We have observed that the distribution $P_n(s)$ appears to be highly model-dependent. This makes it difficult to translate information about the exponential spreading of the wavefunction in the semi-infinite chain back to physical space.

	\begin{figure}
		\centering
		\includegraphics{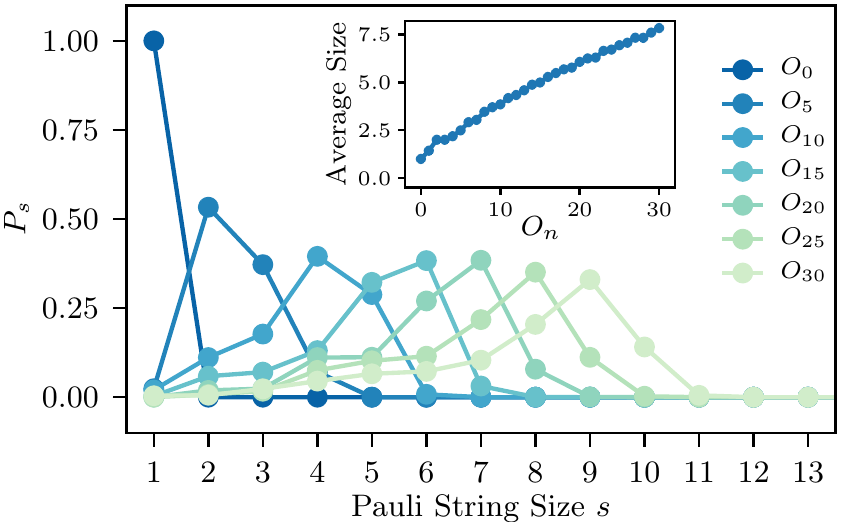}
		\caption{The size distribution of the Pauli strings in the Krylov vectors $\O_n$ for the Hamiltonian $H_1$ with parameters and initial operator as in Fig. \ref{fig:linear_lanczos_evidence}. Though the distribution drops quickly after its peak, $P_n(s)$ is supported on $[0,\lfloor n/2 \rfloor + 2]$.}
		\label{fig:position_distribution}
	\end{figure}

	\section{A Family of Exact Solution with Linear Growth}
	\label{app:universal}
	
	This section will provide a derivation for the exact solution~\eqref{eq:universal_wavefunction_main_text} of the 1d quantum mechanics problem with Lanczos coefficients 
	\begin{equation} b_n  = \alpha \sqrt{n(n-1+\eta)} \label{eq:bn_appendix} \,. \end{equation} 
	To solve this problem, notice that our infinite, tri-diagonal matrix is actually quite a familiar setup. If instead we had $b_n = \sqrt{n}$, then $L$ would be the matrix representing the Hamiltonian for the quantum harmonic oscillator in the basis of raising and lowering operators. So really this is just a 1d quantum mechanics problem, albeit not a standard one. In particular, it is known that the system described by $L$ has very high symmetry, due to an infinite-dimensional representation of the Lie algebra $\mathfrak{su}(1,1)$, enabling us to find an exact solution \cite{hodges2007bernoulli, sukumar2007quantum}. Indeed, there is a rich mathematical literature on the close connections between representations of $\mathfrak{su}(1,1)$, the combinatorics of Motzkin paths, and Meixner orthogonal polynomials \cite{hetyei2010meixner,viennottheorie}. Our solution will be a simple application of these mathematical results.
	
	We start with some generalities on orthogonal polynomials. Define $L_{(n)} = L_{0\le i < n, 0\le j < n}$ to be the $ n \times n$ matrix in the upper-left block of $L$. For example, 
	\begin{equation}
	L_{(3)} = \begin{pmatrix}
	0 & b_1 & 0\\
	b_1 & 0 & b_2\\
	0 & b_2 & 0\\
	\end{pmatrix}.
	\end{equation}
	We then define polynomials for each $n$ via
	\begin{equation}
	Q_n(z;\alpha,\eta) = \det\left( z- L_{(n)} \right).
	\label{eq:Q_polynomial_defn}
	\end{equation}
	By performing a cofactor expansion for the determinant on the $n$th row, the $Q$'s admit a three-term recursion relation
	\begin{equation}
	Q_{n+1}(z) = zQ_{n}(z) -  b_{n}^2 Q_{n-1}(z),
	\label{eq:Q_recusion}
	\end{equation}
	together with initial conditions $Q_{0}(z) = 1$ and $Q_{-1}(z) = 0$. Eq.~\eqref{eq:Q_recusion} should be compared with 
	\begin{equation} L e_n = b_{n+1} e_{n+1} + b_n e_{n-1} \,, \label{eq:L_en} \end{equation}
	where $\st{e_n}$ is the natural orthonormal basis of $L$. In fact, \eqref{eq:Q_recusion} and \eqref{eq:L_en} are equivalent, under the identification:
	\begin{equation}
	Q_n(z) = \left[\; \prod_{k=1}^n b_k\right] {e_n} \,,\,  
	z^n = {L}^n { e_0 }\,. \label{eq:dictionary}
	\end{equation}
	Therefore, the polynomials $Q_n(z)$ are orthogonal, but not normalized. Instead they are monic, i.e., the highest order coefficient is unity: $Q_n(z) = z^n + \BigO(z^{n-1})$.
	
	By construction, both $\st{Q_k(z)}$ and $\st{z^n}$ are a  basis of $\C[z]$ and can be related by a triangular linear transform with matrix elements $\mu_{n,k}$:
	\begin{equation}
	z^n = \sum_{k=0}^n \mu_{n,k} Q_k(z)  \,.
	\end{equation}
	Combined with \eqref{eq:dictionary}, and by orthonormality of $\st{e_n}$, we have
	\begin{equation}
	\pbraket{ e_d \vert  L^n \vert  e_0  } =  \mu_{n,d} \prod_{k=1}^d b_k    \,,
	\end{equation}
	and therefore
	\begin{equation} \label{eq:general_mu}
	\pbraket{ e_d \vert  e^{iLt} \vert  e_0  } = 
	\prod_{k=1}^d b_k  \sum_{n = 0}^{\infty} \frac{(it)^{n}}{n!} \mu_{n,d} \,.
	\end{equation}
	The statements so far are general and apply to any set of Lanczos coefficients.
	
	In the specific case $b_n = \sqrt{n (n-1+\eta)}$ (the extra overall factor $\alpha$ in \eqref{eq:bn_appendix} can be recovered by a simple time rescaling), one may recognize from the recursion relation \eqref{eq:Q_recusion} that $Q_n$'s are a special case of the Meixner polynomials of the second kind~\cite{OEIS_meixner}. They are a non-classical family of orthogonal polynomials defined by the following three-term recursion:~\cite{ismail2009classical,koekoek2010hypergeometric}
	\begin{align}
	M_{n+1}(z;\delta,\eta) \ &=\ (z-\lambda_n) M_{n}(z;\delta,\eta) - b_{n}^2 M_{n-1}(z), \nonumber\\
	\lambda_n \ &=\ (2n+\eta)\delta, \label{eq:Meixner_poly}  \\
	b_n^2 \ &=\ \left( \delta^2 +1 \right)n (n-1+\eta).  \nonumber
	\end{align}
	In particular, $Q_n(z) = M_n(z; \delta=0, \eta)$.
	For these polynomials, the matrix elements $\mu_{n,d}$ have been exactly calculated, in terms of the following generating function~\cite{viennottheorie}:
	\begin{align}
	& \sum_{n=0}^\infty \sum_{d=0}^n \mu_{n,d} w^d  \frac{\tau^n}{n!} 
	\nonumber
	\\  = \,& \frac{\sec(\tau)^\eta}{ (1-\delta \tan(\tau))^{\eta}} \exp\left( w \frac{\tan(\tau)}{1-\delta \tan(\tau)} \right). 	\label{eq:inverse_Meixner_generating_fcn}
	\end{align}
	As a side note, we mention that the above generating function, referred to as that of the ``inverse polynomials'' in the theory of orthogonal polynomial, is closely related to the generating function of Meixner polynomials themselves. The latter has also a closed form expression, known to be of Sheffer type~\cite{hetyei2010meixner,ismail2009classical}:
	\begin{align}
	&\sum_{n \ge 0} M_n(z;\delta,\eta) \frac{\tau^n}{n!}\\
	\nonumber
	&\hspace{2em}=\left[ (1+ \tau \delta)^2 + \tau^2 \right]^{-\eta/2} \exp\left(z \arctan \left( \frac{\tau}{1+ \tau \delta} \right)  \right).
	\end{align}
	
	Now, taking $\delta=0$ and the series coefficient of $w^d$ in \eqref{eq:inverse_Meixner_generating_fcn}, we have 
	$$ \sum_{n=0}^\infty \mu_{n,d} \frac{ \tau^n}{n!} = \frac1{d!} \sec(\tau)^\eta \tan(\tau)^d  \,.  $$
	Applying this to \eqref{eq:general_mu}, and recalling $b_n=\sqrt{n(n-1+\eta)}$, we obtain the wavefunction solution
	\begin{equation}
	\pbraket{ e_n \vert  e^{iLt} \vert  e_0  } = i^n \sqrt{\frac{(\eta)_n}{n!}} \tanh( t)^n \sech( t)^\eta,
	\label{eq:wavefunction_Meixner}
	\end{equation}
	where $(\eta)_n = \eta(\eta+1)\cdots (\eta + n-1)$ is the Pochhammer symbol.  The general solution for $b_n = \alpha \sqrt{n(n-1+\eta)}$ can be obtained by a simple rescaling $t \mapsto \alpha t$, and is precisely Eq. \eqref{eq:universal_wavefunction_main_text} of the main text where, of course,
	$\pbraket{ \O_n \vert  e^{i\mathcal{L}t} \vert  \O_0  }  = \pbraket{ e_n \vert  e^{iLt} \vert  e_0  }$. The special case $\eta = 1$ of this family of solutions is well-known~\cite{viswanath2008recursion,lee2001ergodic}. To the best of our knowledge, the general solution \eqref{eq:wavefunction_Meixner} has not been applied to the recursion method.

	\section{Derivation of the q-Complexity Bound}
	\label{app:bound}
	This Appendix will derive Eq. \eqref{eq:Qtnt}, $\pbraket{\mathcal{Q}}_t \le C \pbraket{n}_t$ for $C = 2M$. The main idea of is that the definition of $\mathcal{Q}$ guarantees that the eigenbasis of $\mathcal{Q}$ is dilated by a factor of at most $C$ compared to the Krylov basis.

	We first show that the Krylov basis vectors have a bounded number of components in the $\mathcal{Q}$ basis due to the dilation property.  For any operator $\Phi$ where there is an $R > 0$ such that $\pbraket{q_a|\Phi} = 0$ for $q_a > R$, the hypothesis \eqref{eq:L_and_q} implies that $\pbraket{q_a|\L|\Phi} = 0$ for $q_a > R+M$. Using \eqref{eq:O_and_q},  as a base case for induction, we have $\pbraket{q_a|\L^n|\O} = 0$ for $q_a > M(n+1)$ and, in particular, for $q_a > C n$. By the construction of the Krylov basis,
	\begin{equation}
	\pbraket{q_a|\O_n} = 0 \quad \text{if } q_a > Cn.
	\label{eq:On_Q_bound}
	\end{equation}
	We claim that \eqref{eq:On_Q_bound} implies 
	\begin{equation} \pbraket{ \Phi \vert  \mathcal{Q} \vert \Phi } \leq C \pbraket{ \Phi \vert  n \vert \Phi } 
	\label{eq:Qngeneral} 
	\end{equation}
	for any operator wavefunction $\Phi$; taking $\Phi = \O(t)$, we obtain \eqref{eq:Qtnt}.
	
	To show \eqref{eq:Qngeneral}, we introduce projectors to large spectral values in the Krylov and $\mathcal{Q}$ bases, respectively:
	\begin{equation}
	\mathcal{P}_n^K = \sum_{m\ge n} \pket{\O_m} \pbra{\O_m}, \quad
	\mathcal{P}_q^Q = \sum_{a\; :\; q_a\ge q} \pket{q_a} \pbra{q_a}.
	\label{eq:Krylov_Q_projectors}
	\end{equation}
	Then, we have for $n = q/C$,
	\begin{equation*}
	\mathcal{P}_q^Q  (1 - \mathcal{P}_{n = q/c}^K )   = 
	\sum_{a\; :\; q_a\ge q}   \sum_{m < n}   \pket{q_a}  \pbraket{q_a \vert O_m } \pbra{\O_m} = 0 ,
	\end{equation*}
	because $m < n = q/C \le q_a / C$,  $\pbraket{q_a \vert O_m } = 0 $ by \eqref{eq:On_Q_bound}. Equivalently,
	\begin{equation}
	\mathcal{P}_q^Q \mathcal{P}_{q/c}^K  = \mathcal{P}_q^Q \,. \label{eq:image1}
	\end{equation}
	
	Applying this equation and its Hermitian conjugate, we have
	\begin{equation}
	\begin{aligned}
	\pbraket{\Phi|\mathcal{P}^Q_q|\Phi}
	\ &=\
	\pbraket{\Phi| \mathcal{P}^Q_q \mathcal{P}^K_{q/C} |\Phi}\\
	\ &=\
	\pbraket{\Phi|\mathcal{P}^K_{q/C} \mathcal{P}^Q_q \mathcal{P}^K_{q/C}|\Phi}\\
	\ &\le \ \pbraket{\Phi|\mathcal{P}^K_{q/C}\mathcal{P}^K_{q/C}|\Phi}\\
	\ &=\ \pbraket{\Phi|\mathcal{P}^K_{q/C}|\Phi}. \label{eq:projector_inequality}
	\end{aligned}
	\end{equation}
	where the inequality follows from the fact that $\mathcal{P}^Q_q$ is a projector.
	Finally we need a standard integration-by-parts identity that converts the expectation value to an integral over the projectors:
	\begin{equation}
	\begin{aligned}
	&	\pbraket{\Phi| \mathcal{Q}^k |\Phi} =  \int_0^\infty dq \; k q^{k-1} \pbraket{\Phi|\mathcal{P}_q^Q|\Phi} \,,\,   \\ 
	&	\pbraket{\Phi| n^k |\Phi} =  \int_0^\infty d n \;  k  n^{k-1}  \pbraket{\Phi|\mathcal{P}_n^K|\Phi} 
	\end{aligned}
	\end{equation}
	for any $k = 1, 2, 3, \dots$. Combining the case $k = 1$ and \eqref{eq:projector_inequality}, we obtain 
	\begin{equation}
	\begin{aligned}
	\pbraket{\Phi|\mathcal{Q}|\Phi} 
	\ &=\ \int_0^\infty dq \; \pbraket{\Phi|\mathcal{P}_q^Q|\Phi}\\
	\ &\le \ \int_0^\infty dq \; \pbraket{\Phi|\mathcal{P}_{q/C}^K|\Phi}\\
	\ &=\ C \pbraket{\Phi|n|\Phi} \,,
	\end{aligned}
	\end{equation}
	which finishes the proof. More generally, for any $k$, we have
	\begin{equation}
	\pbraket{\mathcal{Q}^k}_t \le  C^k \pbraket{n^k}_t \,. \label{eq:Qktnkt}
	\end{equation}
	This is useful as a bound on the growth rate of higher moments of the q-complexity super-operator. See Section~\ref{sec:classical} for an application.

	\section{Geometric Origin of the Upper Bounds}	\label{app:geometric_bound_1d}

	In this appendix we derive the geometric upper bound for the Lanczos coefficients in one-dimensional quantum systems. 
	The main object of our analysis will be the growth of the moments $\mu_{2n} = \pbraket{\O|\L^{2n}|\O} = \dn{\L^n \O}^2$. Moments and Lanczos coefficients are equivalent, and Appendix \ref{app:recursion-review} details how to translate between them. 
	
    To warm up, we first show a bound corresponding to linear growth (using essentially the same argument as in ~\cite{abanin15,abanin17}). This is asymptotically tight in $d>1$. Suppose we have a $2$-local Hamiltonian $H = \sum_x h_x$ and a $1$-local operator $\O$ (the general case of $r$-local $h_x$ and $r$-local $\O$ can be reduced to the previous case by a block renormalization step that groups consecutive sites into renormalized sites). The Liouvillian becomes a sum of terms $\mathcal{L} = \sum_x \ell_x$ with $\ell_x = [h_x,\cdot]$. We suppose that the local terms are uniformly bounded, i.e., for all $x$, $\dn{h_x} \le \mathcal{E}$.  Now, the moment $\mu_{2n}$ is the norm-squared of the sum
	\begin{equation}
	\L^n \O = \sum_{x_1,x_2, \dots,x_n} \ell_{x_n} \cdots \ell_{x_2} \ell_{x_1} \O. 
	\label{eq:path_integral_moments}
	\end{equation}
	This sum is highly constrained by the spatial structure of the spin chain. The operator $\O$ is supported only on one site, and the applications of the Liouvillian grow that support at the edges. Each term in \eqref{eq:path_integral_moments} can be visualized as a discrete quantum circuit, where each gate $\ell_{x_{k+1}}$ must act on at least one site that is already in the support of $\ell_{x_k}\cdots \ell_{x_1}\O$ --- otherwise the term vanishes due to the commutator. This condition is satisfied by at most $(k+1) \le 2 k$ positions $x_k$, so the total number of non-zero terms in \eqref{eq:path_integral_moments} is at most $2^n n!$ for large $n$. The value of each non-zero term is itself bounded due to the finite local bandwidth $\mathcal{E}$, so
	$\dn{\ell_{x_n}\cdots \ell_{x_1}\O}^2 \le (2\mathcal{E})^{2n}$. By the triangle inequality, we have
	\begin{equation}
	\mu_{2n} = \dn{\L^n \O}^2 \le (n!)^2 (4 \mathcal{E})^{2n}.
	\label{eq:upper_bound_moments}
	\end{equation} 
	By Stirling's formula, the right hand side has the same asymptotics as \eqref{eq:moment_growth}, which corresponds to linear growth of the $b_n$'s. Hence \eqref{eq:upper_bound_moments} implies that the Lanczos coefficients can grow at most linearly in any dimension.
	
	Notice that, the bound comes essentially from counting the number of sequences $x_1, \dots, x_n$ that give rise to a nonzero contribution to \eqref{eq:path_integral_moments}. In what follows we show that, in one dimension, there is a sharper upper bound on this number, leading to the sub-linear growth announced in Section~\ref{sec:1dspecial}. 
	For this, we suppose without loss of generality that $\O$ is supported on site $0$ and $h_x$ on sites $x$ and $x+1$. Then it is not hard to see that a $ \ell_{x_n} \cdots \ell_{x_2} \ell_{x_1} \O \neq 0$ \textit{only if} for all $k = 1, \dots, n$,
	\begin{align}
	    &L_k \le  x_k \le R_k   \,, \, \text{ where } \label{eq:admissible} \\
	  &L_k :=  \min\{x_1, \dots, x_{k-1}, 0 \} - 1  \,,\, \nonumber \\
	  &R_k := \max\{x_1,\dots,x_{k-1}, -1\} + 1  \,. \nonumber
	\end{align}
	We define $\mathcal{P}_n$ to be the set of $(x_1, \dots, x_n)$'s that satisfy \eqref{eq:admissible} and denote its size by $P_n := |\mathcal{P}_n|$. Then, similarly to \eqref{eq:upper_bound_moments}, we have
	\begin{equation}
	    \mu_{2n} \le P_n^2 (2 \mathcal{E})^{2n} \,. \label{eq:path_upper_bound}
	\end{equation}
    Hence bounding $\mu_{2n}$ reduces to bounding $P_n$, which is a completely combinatorial problem. 
    
    To produce this combinatorial bound, we partition the set $\mathcal{P}_n$ as follows 
    \begin{align}
       & \mathcal{P}_n = \bigcup_{\ell=1}^{n}\mathcal{P}_{n,\ell} \,,\, \text{ where }  \nonumber \\
    &    \mathcal{P}_{n,\ell} :=  \{(x_1, \dots, x_n) \in \mathcal{P}_n: 
        \ell = L_n - R_n \} \,. \label{eq:partition}
    \end{align}
   Intuitively, if the support of the operator grows to size $\ell + 1$ after $n$ applications of Liouvillian, then $(x_1, \dots, x_n) \in  \mathcal{P}_{n,\ell}$. By ``size", we mean the distance between the endpoints, disregarding the ``holes'' between them. In the 1d case, the operator size can only grow in two places: the left and right sides. Therefore, for any $(x_1, \dots, x_n) \in  \mathcal{P}_{n,\ell}$, $x_k = L_k$ or $x_k = R_k$ must hold for $\ell$ values of $k$ among $1, \dots, n$: for each of such $k$'s, one has only two choices for $x_k$. For the remaining $n-\ell$, there are (at most) $\ell$ choices (by \eqref{eq:admissible}, minus 2 boundary choices). Summarizing, we have 
   \begin{equation}
       |\mathcal{P}_{n,\ell}| \le \binom{n}{\ell} 2^{\ell} \ell^{n-\ell} 
       \le 4^n \ell^{n-\ell} \,,
   \end{equation}
   where the binomial coefficient counts the choices of the $\ell$ values. Combining this with \eqref{eq:partition}, we have 
   \begin{equation}
       P_n \le n  4^n  \max_{\ell\in [0, n]} \ell^{n-\ell}  \,.
   \end{equation}
In the limit $n \gg 1$, the maximum is attained at $\ell = n/W(n)$ where $W$ is the product-log function defined by $z = W(z e^z)$. For large $n$, $W(n) = \ln n - \ln \ln n + o(1)$, so
\begin{align}
	P_n \le n 4^n \left( \frac{n}{W(n)} \right)^{n-\frac{n}{W(n)}}
%	= \exp\left[ n \ln n - n \ln \ln n - n + o(n) \right]
	= \frac{n! 4^n}{(\ln n)^n} e^{o(n)}.
\end{align}
where we used $n/W(n) = e^{W(n)}$ and Stirling's formula. Therefore
\begin{equation}
	\mu_{2n}  \le (4 \mathcal{E})^{2n} \frac{(n!)^2}{(\ln n)^{2n}} e^{o(n)} \,,
	\label{eq:moments_upper_bound}
\end{equation}
which grows more slowly than the moment asymptotics corresponding to a linear growth with rate $\alpha$ \eqref{eq:moment_growth1}, $B_n \ll \left( \frac{4 n \alpha}{e \pi} \right)^{2n}$, for \textit{any} $\alpha > 0$. So the Lanczos coefficients corresponding to \eqref{eq:moments_upper_bound} must be sub-linear. 

What, then, is the fastest possible growth of the $b_n$'s in 1D? Although we cannot bound the individual Lanczos coefficients in a useful way from the bound on the moments, we can use the bound on their cumulative product $ \ln \prod_{k=1}^n b_k^2 \le \ln \mu_{2n}$ \eqref{eq:moment_bn_bound} and differentiate with respect to $n$. As a result, we find
\begin{equation}
	b_n = A \frac{n}{W(n)} = A e^{W(n)} \sim{} \frac{A n}{\ln n} \,.
\end{equation}
The bound \eqref{eq:moment_bn_bound} (together with \eqref{eq:moments_upper_bound}) is satisfied asymptotically by the above choice of $b_n$ if and only if $A \le 4 \mathcal{E} / e$. Therefore, $b_n = a e^{W(n)}$ captures the correct asymptotic behavior of the upper-bound in the moments, and qualifies as the maximal growth rate of Lanczos coefficients in 1d.
\end{document}